\documentclass{article}
\usepackage{epsfig}

\date{\today}

\usepackage[intlimits,tbtags]{amsmath} 
\usepackage{amsfonts,amssymb}
\usepackage{amsthm}
\usepackage{amscd}
\usepackage{subfigure}

\begin{document}
\title{Geodesic motion in the space--time of a cosmic string}
\author{{\large Betti Hartmann \footnote{email: b.hartmann@jacobs-university.de} } and 
{\large Parinya Sirimachan \footnote{email: p.sirimachan@jacobs-university.de}}
\\ \\
{\small School of Engineering and Science, Jacobs University Bremen, 28759 Bremen, Germany}  }

\date{}
\newcommand{\dd}{\mbox{d}}
\newcommand{\tr}{\mbox{tr}}
\newcommand{\la}{\lambda}
\newcommand{\ka}{\kappa}
\newcommand{\f}{\phi}
\newcommand{\vf}{\varphi}
\newcommand{\F}{\Phi}
\newcommand{\al}{\alpha}
\newcommand{\ga}{\gamma}
\newcommand{\de}{\delta}
\newcommand{\si}{\sigma}
\newcommand{\bomega}{\mbox{\boldmath $\omega$}}
\newcommand{\bsi}{\mbox{\boldmath $\sigma$}}
\newcommand{\bchi}{\mbox{\boldmath $\chi$}}
\newcommand{\bal}{\mbox{\boldmath $\alpha$}}
\newcommand{\bpsi}{\mbox{\boldmath $\psi$}}
\newcommand{\brho}{\mbox{\boldmath $\varrho$}}
\newcommand{\beps}{\mbox{\boldmath $\varepsilon$}}
\newcommand{\bxi}{\mbox{\boldmath $\xi$}}
\newcommand{\bbeta}{\mbox{\boldmath $\beta$}}
\newcommand{\ee}{\end{equation}}
\newcommand{\eea}{\end{eqnarray}}
\newcommand{\be}{\begin{equation}}
\newcommand{\bea}{\begin{eqnarray}}

\newcommand{\ii}{\mbox{i}}
\newcommand{\e}{\mbox{e}}
\newcommand{\pa}{\partial}
\newcommand{\Om}{\Omega}
\newcommand{\vep}{\varepsilon}
\newcommand{\bfph}{{\bf \phi}}
\newcommand{\lm}{\lambda}
\def\theequation{\arabic{equation}}
\renewcommand{\thefootnote}{\fnsymbol{footnote}}
\newcommand{\re}[1]{(\ref{#1})}
\newcommand{\R}{{\rm I \hspace{-0.52ex} R}}
\newcommand{\N}{{\sf N\hspace*{-1.0ex}\rule{0.15ex}%
{1.3ex}\hspace*{1.0ex}}}
\newcommand{\Q}{{\sf Q\hspace*{-1.1ex}\rule{0.15ex}%
{1.5ex}\hspace*{1.1ex}}}
\newcommand{\C}{{\sf C\hspace*{-0.9ex}\rule{0.15ex}%
{1.3ex}\hspace*{0.9ex}}}
\newcommand{\eins}{1\hspace{-0.56ex}{\rm I}}
\renewcommand{\thefootnote}{\arabic{footnote}}

\maketitle

\bigskip

\begin{abstract}
We study the geodesic equation in the space--time of an Abelian--Higgs string and discuss the motion
of massless and massive test particles. The geodesics can be classified
according to the particles energy, angular momentum and linear momentum
along the string axis. We observe that bound orbits of massive particles are only possible
if the Higgs boson mass is smaller than the gauge boson mass, while massless particles
always move on escape orbits. Moreover, neither massive nor massless  particles can ever reach the string axis for non--vanishing angular momentum. 
We also discuss the dependence of light deflection by a cosmic string as well as the perihelion shift of bound orbits of massive particles on the
ratio between Higgs and gauge boson mass and the ratio between symmetry breaking scale
and Planck mass, respectively.

\end{abstract}
\medskip
\medskip

\section{Introduction}
Cosmic strings have gained a lot of renewed interest over the past years
due to their possible connection to string theory \cite{polchinski}.
These are topological defects \cite{vs} that could have formed in one of the numerous phase transitions in the early universe due to the Kibble mechanism.
Inflationary models resulting from string theory (e.g. brane inflation)
predict the formation of cosmic string networks at the end of inflation \cite{braneinflation}.
It would be very interesting to observe these objects in the universe. 
In recent years the detection of these objects has focused on the Cosmic Microwave background (CMB) data \cite{cmb_cosmic}, though other observational effects such as gravitational
lensing have also been discussed \cite{khol}. 
Here we discuss the possibility that cosmic strings might be detected due to
the way that test particles move in their space--time.

Until now the geodesic motion of test particles in space--times containing 
cosmic strings has been mostly studied in the limit of vanishing width of the cosmic string. Since the space--time of an infinitely thin cosmic string is locally flat \cite{vs} geodesics
are just straight lines. In \cite{gibbons} a general cosmic space--time has been
studied and it has been shown that the geodesics of massless particles must move to infinity
in both direction. Geodesics in cosmic string space--times have also been used
to explain the motion of particles in elastic solids \cite{ppm}. Furthermore, black hole
space--times containing infinitely thin cosmic strings have been investigated
both for static black holes \cite{ag,gm,cb,hhls1} as well as for rotating black holes \cite{gm,
Ozdemir2003,Ozdemir2004,Fernandes2006,hhls2}. In this case, analytic solutions to the geodesic
equation in terms of elliptic functions are possible.

In this paper we are aiming at understanding particle motion in the space--time of a finite
width cosmic string. The underlying field theoretical model is the U(1) Abelian--Higgs model
which has string--like solutions \cite{no}. 
The gravitational properties of Abelian--Higgs strings have also been studied
in detail by minimally coupling the Abelian--Higgs model to gravity. Far away from the 
core of the string, the space-time has a deficit angle, i.e. corresponds to
Minkowski space-time minus a wedge \cite{garfinkle}. The deficit angle is in linear
order proportional to the energy per unit length of the string. 
If the vacuum expectation value (vev) of the Higgs field is sufficiently large
(corresponding to very heavy strings that have formed at energies much bigger than the GUT scale), the deficit angle becomes larger than $2\pi$. These solutions
are the so-called ``supermassive strings'' studied in \cite{gl} and possess
a singularity at a maximal value of the radial coordinate, at which the
angular part of the metric vanishes. Interestingly, it was realized
only a few years ago \cite{clv,bl} that both the globally regular string solution
as well as the supermassive solution have ``shadow'' solutions that exist for the same parameter values. The string-like solutions with deficit angle $ < 2\pi$ have a
shadow solution in the form of so-called Melvin solutions, which have a different
asymptotic behaviour of the metric and have higher energies than their string counterparts (and are thus very likely cosmologically not relevant). The supermassive
string solutions on the other hand have shadow solutions of Kasner-type.
Kasner solutions possess also a singularity at some finite radial distance, the difference is that for Kasner solutions, the $tt$-component of the metric vanishes at this maximal radial distance, while the angular part of the metric diverges there.
Since both supermassive as well as Kasner solutions contain space-time singularities
they are surely of limited interest for cosmological applications. In this paper
we thus concentrate on the motion of test particles in the space--time of a 
gravitating cosmic string that possesses a deficit angle $ < 2\pi$. Since the field theoretical
solutions and in particular the metric functions can only be determined numerically analytic
results are not possible for the geodesic equation. This is similar to the geodesic
motion in the space--time of a gravitating magnetic monopole \cite{kkl}, where also only
numerical results are possible. 

Our paper is organised as follows: in Section 2, we discuss the field theoretical
model that possesses string--like solutions and we also give the geodesic equation.
In Section 3 we discuss our numerical results, in particular we give examples
of orbits and demonstrate how the ratio between the symmetry breaking scale as well as the ratio
between the Higgs and gauge boson mass influence our results.
We conclude in Section 4.

\section{The model}
In the following we will first present the space--time of an Abelian--Higgs string
and then give the geodesic equation that describes the motion of massless and massive
test particles in the space--time of an Abelian--Higgs string.

\subsection{The space--time of an Abelian--Higgs string}
The Abelian--Higgs model is a field theoretical model with cosmic string solutions \cite{no}.
Here, we couple this model minimally to gravity. The action then reads
\begin{equation}
\label{action}
S=\int d^4 x \sqrt{-g} \left( \frac{1}{16\pi G} R + {\cal L}_{m} \right)
\end{equation}
where $R$ is the Ricci scalar and $G$ denotes Newton's constant. The matter Lagrangian
${\cal L}_{m}$ is given by
\begin{equation}
{\cal L}_{m}=D_{\mu} \phi (D^{\mu} \phi)^*-\frac{1}{4} F_{\mu\nu} F^{\mu\nu}
-\frac{\lambda}{4}\left(\phi\phi^*-\eta^2\right)^2
\end{equation} 
with the covariant derivative $D_\mu\phi=\nabla_{\mu}\phi-ie A_{\mu}\phi$
and the
field strength tensor $F_{\mu\nu}=\partial_\mu A_\nu-\partial_\nu A_\mu$
of the U(1) gauge potential $A_{\mu}$ with coupling constant $e$. The field 
$\phi$  is a complex scalar field (Higgs field).

The most general, cylindrically symmetric line element invariant under boosts
along the $z-$direction is
\begin{equation}
\label{metric}
ds^2=N^2(\rho)dt^2-d\rho^2-L^2(\rho)d\varphi^2-N^2(\rho)dz^2 \ .
\end{equation}
For the matter and gauge fields, we have \cite{no}:
\begin{equation}
\phi(\rho,\varphi)=\eta h(\rho)e^{i n\varphi} \ \ , \ \ 
A_{\mu}dx^{\mu}=\frac {1}{e}(n-P(\rho)) d\varphi \ , 
\end{equation}
where $n$ is an integer indexing the vorticity of the Higgs field  around the $z-$axis.
As is apparent from the Ansatz the cosmic string possesses a magnetic field
along the $z$--axis with \cite{clv}
\begin{equation}
 B_z(\rho)=-\frac{1}{eL(\rho)}\frac{dP(\rho)}{d\rho}  \ .
\end{equation}

There are three mass scales in our model: the Higgs mass $M_{\rm H}=\sqrt{\lambda} \eta$,
the gauge boson mass $M_{\rm W}=\sqrt{2}e\eta$ as well as the Planck mass $M_{\rm Pl}=G^{-1/2}$.
The Bogomolny limit $\lambda=2e^2$ \cite{bps} corresponds to equal gauge and Higgs boson mass.
The cosmic string has finite width. In fact there is a scalar core with width $\rho_{\rm H}\sim M_{\rm H}^{-1}$
and a magnetic flux tube core with width $\rho_{\rm W}\sim M_{\rm W}^{-1}$. In the Bogomolny limit
$\rho_{\rm H}=\rho_{\rm W}$.

We can then do the following rescaling
\begin{equation}
\label{scaling}
\rho\rightarrow \frac{\rho}{e\eta} \ \ \ , \ \ \ L\rightarrow \frac{L}{e\eta}\ .
\end{equation}
such that the total Lagrangian only depends on the following dimensionless coupling constants
\begin{equation}
\gamma=8\pi G\eta^2 \ \ , \  \  
\beta=\frac{\lambda}{e^2} \ .
\end{equation}
Note that $\gamma$ is proportional to the square of the ratio of the symmetry
breaking scale $\eta$ and the Planck mass $M_{\rm Pl}$, while $\beta$ is proportional to the
square of the ratio between the  Higgs boson mass $M_{\rm H}$ and
the gauge boson mass $M_{\rm W}$.

Varying the action with respect to the matter fields and metric functions, we
obtain a system of four non-linear differential equations. The Euler-Lagrange equations for the matter field functions read:
\begin{equation}
\label{eqh}
\frac{(N^2Lh')'}{N^2L}=\frac{P^2 h}{L^2}+\frac{\beta}{2} h(h^2-1)
\end{equation}
\begin{equation}
\frac{L}{N^2}\left(\frac{N^2P'}{L}\right)'=2 h^2 P \ ,
\end{equation}
while the Einstein equations are
\begin{eqnarray}
\label{N1}
\frac{(LNN')'}{N^2 L}&=& \gamma\left[\frac{(P')^2}
{2 L^2}
-  \frac{\beta}{4}\left(h^2-1\right)^2   \right] 
\end{eqnarray}
and
\begin{eqnarray}
\label{N2}
\frac{(N^2L')'}{N^2L}&=&-\gamma\left[\frac{2 h^2 P^2}
{L^2}+\frac{(P')^2}{2 L^2} + \frac{\beta}{4}\left(h^2-1\right)^2 \right]  \ .
\end{eqnarray}
Here and in the following the prime denotes the derivative with respect to $\rho$.

The set of differential equations can only be solved numerically subject to a set of boundary
conditions. 
The requirement of regularity at the origin leads to the  following boundary 
conditions
\begin{equation}
h(0)=0 \  \ , \ P(0)=n \ , \ N(0)=1, \ N'(0)=0, \ L(0)=0 \ , \ L'(0)=1 \ ,
\label{eom1}
\end{equation}
while 
the finiteness of the energy per unit length requires
\begin{equation}
h(\infty)=1 \ , \ P(\infty)=0 \   \ .
\end{equation}

In this paper we are interested in test particle motion in the space--time of a cosmic
string with deficit angle $\Delta < 2\pi$, i.e. we do not consider supermassive
strings or space--times of Melvin-- or Kasner--type in this paper. The metric functions then have the following behaviour at infinity
\begin{equation}
\label{asymptotics}
 N(\rho \gg 1)=c_1 \ \ , \ \ L(\rho\gg 1)=c_2\rho + c_3
\end{equation}
where $c_1$, $c_2$ and $c_3$ are constants that depend on $\beta$ and $\gamma$ \cite{bl}.
We have $c_1=1$ independent on $\gamma$ and $c_2=1-2\gamma$ in the Bogomolny limit $\beta=2$.
$c_1$ and $c_2$ both
decrease for fixed $\gamma$ and increasing $\beta$ with $c_1 > 1$, $c_2 > 1-2\gamma$ for $\beta < 2$,
and $c_1 < 1$, $c_2 < 1-2\gamma$ for $\beta > 2$. Moreover,
$c_1$ increases with $\gamma$ for fixed $\beta <2$, while $c_1$ decreases with $\gamma$ for fixed
$\beta > 2$. 

In the Bogomolny limit $\beta=2$ which corresponds to $M_{\rm H}=M_{\rm W}$ we have
that $P'=L(h^2-1)$ \cite{bps}. Inserting this into the Einstein equation (\ref{N1}) and using the
boundary conditions it is easy to see that in this case $N(\rho)\equiv 1$, i.e. the space--time is
only determined by $L(\rho)$. 

We define as inertial mass per unit length of the solution
\begin{equation}
 \mu=\int \sqrt{-g_3} T^0_0 d\rho d\varphi
\end{equation}
where $g_3$ is the determinant of the $2+1$-dimensional space-time given by $(t,\rho,\varphi)$.
This then reads:
\begin{equation}
 \mu=2\pi\int_{0}^{\infty} NL \left((h')^2 + \frac{(P')^2}{2 L^2} +  \frac{h^2 P^2}{L^2}+ 
\frac{\beta}{4}\left(h^2-1\right)^2\right) d\rho
\end{equation}
In flat space--time $\gamma=0$ and in the Bogomolny limit $\beta=2$ the energy per unit length is directly proportional
to the winding number $n$ and is given by $\mu=2\pi n$ \cite{bps}, while the general expression for
$\mu$ in flat space--time is
\begin{equation}
 \mu=2\pi n f(\beta)
\end{equation}
where $f$ is a slowly varying function of $\beta$ which is $f< 1$ for $\beta < 2$, $f=1$ for $\beta=2$ and
$f > 1$ for $\beta > 2$. For $\gamma > 0$ and fixed $\beta$ the mass per unit length of the solution decreases with increasing $\gamma$. 

In linear order the deficit angle $\Delta$ 
of the space--time is proportional to the product
of the gravitational coupling and the energy per unit length of the string $\gamma\mu$. Taking the non--linear effects into account $\Delta$ is related to $c_2$ (see (\ref{asymptotics})) by 
\begin{equation}
 \Delta=2\pi(1-c_2) \ .
\end{equation}
Following the discussion above, the deficit angle increases when increasing $\gamma$ and
$\Delta=4\pi \gamma$ for $\beta=2$, while $\Delta < 4\pi \gamma$ for $\beta < 2$ and
$\Delta > 4\pi \gamma$ for $\beta > 2$. Since we are only considering space--times with $\Delta < 2\pi$ in this paper there are restrictions on the parameters $\gamma$ and $\beta$. 
In \cite{bl} the value of $\gamma$ at which $\Delta =2\pi$ has already been investigated
for some particular values of $\beta$. We come back to this in the numerical results section.

\subsection{The geodesic equation}
We consider the geodesic equation
\begin{equation}
 \frac{d^2 x^{\mu}}{d\tau^2} + \Gamma^{\mu}_{\rho\sigma} \frac{dx^{\rho}}{d\tau}\frac{dx^{\sigma}}{d\tau}=0  \ ,
\end{equation}
where $\Gamma^{\mu}_{\rho\sigma}$ denotes the Christoffel symbol given by
\begin{equation}
 \Gamma^{\mu}_{\rho\sigma}=\frac{1}{2}g^{\mu\nu}\left(\partial_{\rho} g_{\sigma\nu}+\partial_{\sigma} g_{\rho\nu}-\partial_{\nu} g_{\rho\sigma}\right)
\end{equation}
and $\tau$ is an affine parameter such that for time--like geodesics $d\tau^2=g_{\mu\nu}dx^{\mu} dx^{\nu}$ corresponds to proper time.

The geodesic Lagrangian $\mathcal{L}_{\rm g}$ for a point particle in the space--time (\ref{metric}) reads
\begin{eqnarray}
\label{lagrangian_geo}
\mathcal{L}_{\rm g}=\frac{1}{2}g_{\mu\nu}\frac{dx^{\mu}}{ds}\frac{dx^{\nu}}{ds}=\frac{1}{2}\varepsilon
=\frac{1}{2}\left[N^2\left(\frac{dt}{d\tau}\right)^{2}-\left(\frac{d\rho}{d\tau}\right)^{2}-L^2
\left(\frac{d\varphi}{d\tau}\right)^{2}-N^2\left(\frac{dz}{d\tau}\right)^{2}\right]
 \ ,  \end{eqnarray}
where $\varepsilon=0$ for massless particles and $\varepsilon=1$ for massive particles, respectively. 

The constants of motion are the energy $E$, the angular
momentum of the particle that is aligned with the axis of the string (here the $z$-axis) $L_z$
and the momentum $p_z$ in $z$--direction
\begin{eqnarray}
\label{constants}
E:=N^2\frac{dt}{d\tau}\ \ , \ \
L_z:=L^2\frac{d\varphi}{d\tau}  \ \ , \ \ 
p_z:=N^2\frac{dz}{d\tau}  \ .
\end{eqnarray}
Using these constants of motion and the rescaling (\ref{scaling}) as well as letting
${\cal E}\rightarrow {\cal E}/(e^2\eta^2)$, $p_z\rightarrow p_z/(e\eta)$, $L_z\rightarrow L_z/(e^2\eta^2)$  we can rewrite (\ref{lagrangian_geo}) as follows
\begin{equation}
\label{eq1}
 \frac{1}{2}\left(\frac{d\rho}{d\tau}\right)^2 = {\cal E} - V_{\rm eff}(\rho)  \ ,
\end{equation}
where ${\cal E}=(E^2-\varepsilon)/2$ and $V_{\rm eff}(\rho)$ is the effective potential
\begin{equation}
\label{potential}
 V_{\rm eff}(\rho)=\frac{1}{2}\left[E^2\left(1-\frac{1}{N^2}\right)+ \frac{p_z^2}{N^2} + \frac{L_z^2}{L^2}\right]  \ .
\end{equation}
As is obvious from (\ref{eq1}) test particle motion is only possible for
${\cal E} - V_{\rm eff}(\rho) > 0$.
The $\rho$-component of the geodesic equation $\frac{d^2\rho}{d\tau^2}=-\frac{dV_{\rm eff}}{d\rho}$ then reads
\begin{equation}
\frac{d^2\rho}{d\tau^2}=\left(p_z^2-E^2\right) \frac{N'}{N^3} + L_z^2 \frac{L'}{L^3} \ .
\end{equation}
For our numerical calculations, we have rewritten the components of the geodesic equation in the following way

\begin{eqnarray}
\label{neweq1}
d\varphi&=&\pm \frac{L_z d\rho}{L(\rho)^2\left(\frac{E^2-p_z^2}{N(\rho)^2}-\frac{L_z^2}{L(\rho)^2}-\varepsilon\right)^{1/2}}  \ , \\
\label{neweq2}
dz&=&\pm \frac{p_z d\rho}{N(\rho)^2 \left(\frac{E^2-p_z^2}{N(\rho)^2}-\frac{L_z^2}{L(\rho)^2}-\varepsilon\right)^{1/2}}  \ , \\
\label{neweq3}
dt&=&\pm \frac{E d\rho}{N(\rho)^2 \left(\frac{E^2-p_z^2}{N(\rho)^2}-\frac{L_z^2}{L(\rho)^2}-\varepsilon\right)^{1/2}}  \ .
\end{eqnarray}
For positive angular momentum $L_z$ the positive and the negative sign denote outward and inward motion, respectively. The motion becomes non-planar for $p_z$ $\neq$ 0.

\section{Numerical results}

The equations of motion (\ref{eqh})-(\ref{N2}) can only be solved numerically. We have done
this using the ordinary differential equation solver COLSYS \cite{colsys}.
The relative errors of the numerical integration are on the order of
$10^{-10}-10^{-13}$. 
The numerical
data for the metric functions $N$ and $L$ then has to be interpolated. This has been
done with a piecewise cubic Hermite interpolating polynomial within MATLAB.
The interpolated data can then be used to find $t(\rho)$, $\varphi(\rho)$
and $z(\rho)$ via (\ref{neweq1})-(\ref{neweq3}).
This latter integration has been done by using the recursive
adaptive Simpson quadrature within MATLAB requiring an absolute error tolerance of $10^{-8}$.
\begin{figure}
\centering
\epsfig{figure=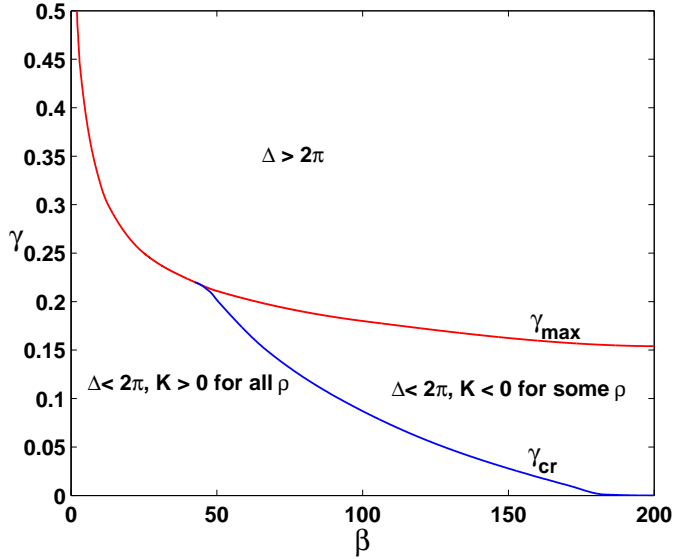,width=10.0 cm}
\caption{\label{gammabeta} We show the value of $\gamma_{\rm max}$, i.e. the value of $\gamma$ at which the deficit angle $\Delta$ is equal
to $2\pi$ in dependence on $\beta$. Below this line cosmic string
solutions with $\Delta < 2\pi$ exist, while above this line the cosmic string
solutions are supermassive with $\Delta > 2\pi$ and hence possess a space--time singularity. 
In this paper we are only interested
in solutions with $\Delta < 2\pi$. We also show $\gamma_{\rm cr}$, i.e. the value of $\gamma$ such that for $\gamma> \gamma_{\rm cr}$ the Gaussian
curvature $K$ of the 2-manifold with coordinates $(\rho,\varphi)$ can have negative values for some $\rho$.  }
\end{figure}

As mentioned above we are only interested in cosmic string space--times with $\Delta < 2\pi$ in this
paper. Since $\Delta$ depends on both $\gamma$ and $\beta$ there are restrictions on these 
parameters. These restrictions have already been investigated in \cite{bl} for
some particular values of $\beta$. Here, we show the domain of existence
of cosmic string solutions with $\Delta < 2\pi$ in Fig.\ref{gammabeta} where we 
give the value of $\gamma_{\rm max}$, i.e. the value of $\gamma$ at which the deficit
angle becomes equal to $2\pi$ in dependence on $\beta$. 

In the following we will distinguish between bound orbits and escape orbits.
Note that when we talk about bound and escape orbits we are referring to the motion
in the $x$--$y$--plane. The particles can, of course, move along the full $z$--axis
from $-\infty$ to $+\infty$ for $p_z\neq 0$.

Bound orbits are orbits on which test particles move 
from a minimal value of $\rho$, $\rho_{\rm min} > 0$ to a maximal value of
$\rho$, $\rho_{\rm max} < \infty$ and back again. These orbits have hence two turning points
with $(d\rho/d\tau)^2=0$. On escape orbits, on the other hand, particles come from $\rho=\infty$,
reach a minimal value of $\rho$, $\rho_{\rm min} > 0$ and move back to $\rho=\infty$, which means that escape
orbits have only one turning point with $(d\rho/d\tau)^2=0$.
Looking at (\ref{eq1}) it is obvious that turning points are located at those $\rho$ at
which  ${\cal E} - V_{\rm eff}(\rho) = 0$.  

For all our numerical calculations we have chosen $n=1$. 

\subsection{The effective potential}

The first observation is that since $N(\rho\rightarrow 0)\rightarrow 1$ and $L(\rho\rightarrow 0)\rightarrow \rho$
there is always an infinite potential barrier at $\rho\rightarrow 0$  for $L_z\neq 0$ which means
that the particles can never reach the $z$-axis, i.e. $\rho=0$. For $L_z=0$ this potential
barrier disappears.

Moreover, in the Bogomolny limit $\beta=2$ with $N(\rho)\equiv 1$ the effective
potential is monotonically decreasing from $V_{\rm eff}(\rho=0)=+\infty$ to $V_{\rm eff}(\rho=\infty)=0$ and ${\cal E} - V_{\rm eff}(\rho) = 0$ can only be fulfilled for
one $\rho$. 
Hence in the Bogomolny limit bound orbits do not exist and test particles can only move on escape orbits. This is different for $\beta\neq 2$ where $N(\rho)$ is non--constant. In fact, all our
numerical results indicate that bound orbits exist only for $\beta <2$ and for massive
particles ($\varepsilon=1$). An indication
that this is correct can be seen when looking at (\ref{eq1}). Bound orbits are possible
whenever $V_{\rm eff}$ possesses at least one extremum, i.e. there is one finite $\rho$ for
which $dV_{\rm eff}/d\rho=0$. Using the explicit form of the effective potential (\ref{potential}) this gives
\begin{equation}
\label{extremum}
 \frac{dV_{\rm eff}}{d\rho}=0 \ \ \Longrightarrow \ \ \frac{N'}{L'}=\frac{L_z^2}{E^2-p_z^2} \frac{N^3}{L^3}    \ .
\end{equation}
Assuming $N$ and $L$ to be positive, the right-hand side of (\ref{extremum}) is always positive.
In addition $L'$ is always positive which can be seen from (\ref{N2}) and the fact that
$L'(0)=0$ and $L'(\infty) > 0$. Now from (\ref{N2}) it is easy to see that 
$L'=0$ for some finite $0 < \rho < \infty$ is excluded. Hence, for extrema to exist we need
to require that $N' > 0$. In fact for $\beta <2$ we find that $N' > 0$, while $N' < 0$ for $\beta > 2$ (see also the results in \cite{clv,bl}). Of course, the requirement
that $N' >0$ does not guarantee that bound orbits exist (this
still depends on the choice of $\mathcal{E}$, $L_z$ and $p_z$), but
for $N' < 0$, i.e. $\beta > 2$ we can exclude the possibility
of bound orbits. Our numerical results demonstrating the change of the effective potential
with $\gamma$ and $\beta$ are shown in Fig.\ref{fignew1}. For fixed $\gamma$ it is obvious that
local extrema exist for $\beta < 2$, while there are no extrema for $\beta \ge 2$. Moreover,
the potential changes only little for changing $\beta > 2$. 

\begin{figure}[h!]
  \begin{center}
    \subfigure[$\gamma=0.3$]{\includegraphics[scale=0.476]{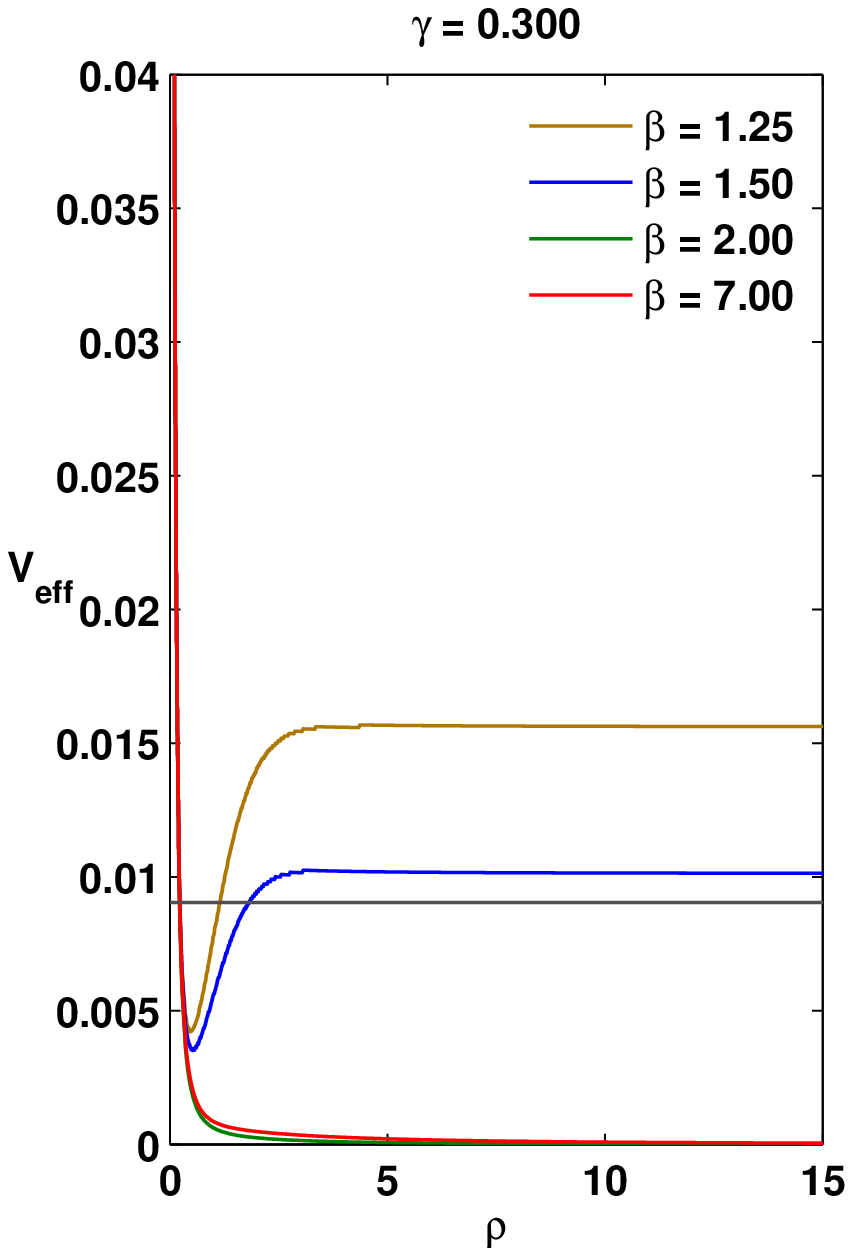}}
    \subfigure[$\beta =1.25$ ]{\label{betaless2}\includegraphics[scale=0.476]{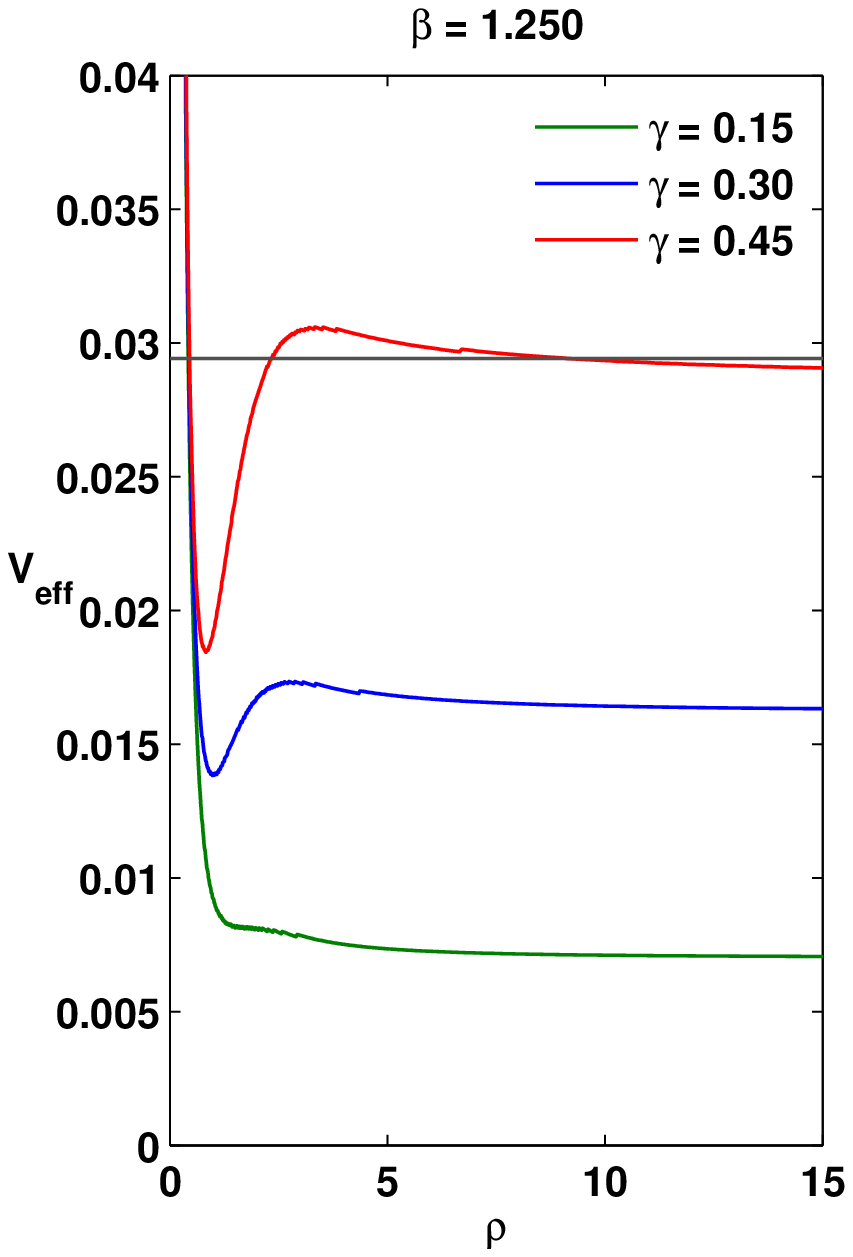}}
    \subfigure[$\beta=2.5$]{\label{betagreater2}\includegraphics[scale=0.476]{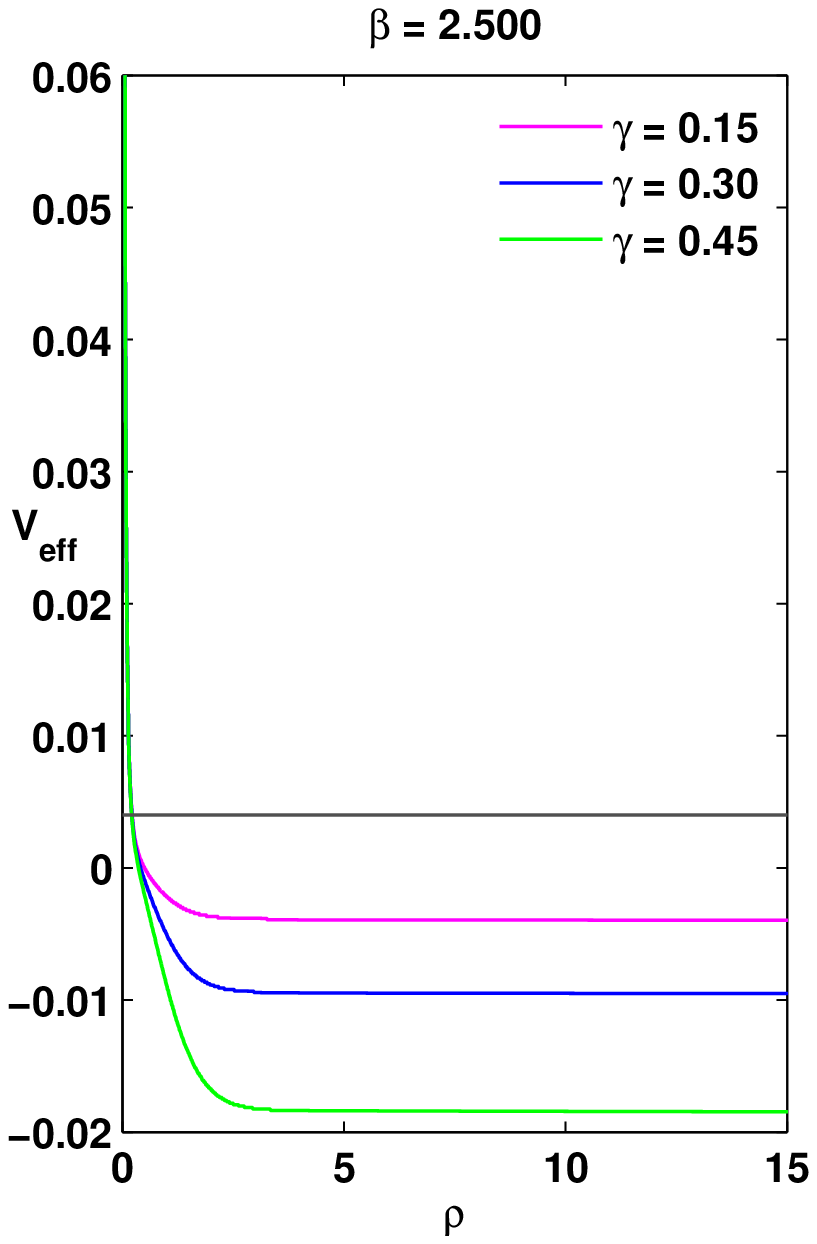}}
    \end{center}
  \caption{The effective potential for the motion of a massive test particle $(\varepsilon=1$)
in the space--time of an Abelian--Higgs string is shown for fixed $\gamma=0.3$, $E=1.009$, $p_z=0$, $L_z=0.03$ and different values
of $\beta$ (a), fixed $\beta=1.25 < 2$, $E=1.029$, $p_z=0$, $L_z=0.1$ and different values of $\gamma$ (b) and
fixed $\beta=2.5 > 2$, $E=1.004$, $p_z=0$, $L_z=0.02$ and different values of $\gamma$ (c). The black horizontal line represents the
value of $\cal{E}$.}\label{fignew1}
  \end{figure}

For fixed $\beta < 2$ (here $\beta=1.25$) local extrema do exist and the value of these extrema
is enhanced with increasing gravitational coupling.  For fixed $\beta > 2$ (here $\beta=2.5$)
no extrema exist.
While the effective potential possesses local extrema for $\beta < 2$ bound orbits exist
only for particular choices of $\mathcal{E}$, $L_z$ and $p_z$. As mentioned above, our results
indicate that bound orbits exist only for massive particles $(\varepsilon=1$).
In order to demonstrate this, we have chosen $\beta=0.5$, $\gamma=0.4$. We show the
effective potential for a massive and a massless test particle in Fig.\ref{fignew2} and
Fig.\ref{fignew3}, respectively. It can be seen from Fig.\ref{fignew2} that
the potential possesses local extrema and that $\mathcal{E}$ intersects the potential twice
for sufficiently low values of $E$ if $\varepsilon=1$. This is different for the massless
case. As can been seen from Fig.\ref{fignew3}, the local extrema of the effective potential
disappear with decreasing $E$ such that when $\mathcal{E}$ is comparable to the asymptotic
value of the effective potential, there are no local extrema at all.
Hence, all our numerical results indicate that bound orbits for massless particles
are excluded. 

In the following we want to compare this result with the analytical result of \cite{gibbons}, which states that for a general cosmic string space--time with topology $\mathbb{R}^2\times \Sigma$
massless test particles must move on geodesics that escape to infinity in both directions, i.e.
closed geodesics are not possible.
The assumption made in \cite{gibbons} is that $\Sigma$ must have positive Gaussian curvature.
To show that $\Sigma$ has positive Gaussian curvature in our case, we rewrite the metric
(\ref{metric}) for massless particles ($ds^2=0$) moving in a plane parallel to the $x$-$y$-plane
as follows
\begin{equation}
 dt^2=\frac{1}{N^2} d\rho^2+ \frac{L^2}{N^2}d\varphi^2 = \tilde{g}_{ij} dx^i dx^j \ \  ,  \ \ i=1,2
\end{equation}
where $\tilde{g}_{ij}$ is the so-called optical metric \cite{acl} of which the spatial
projection of geodesics of massless particles, i.e. light rays are geodesics. 
$\tilde{g}_{ij}$ is the metric of the above mentioned 2-manifold $\Sigma$ and has Gaussian
curvature $K$ given by
\begin{eqnarray}
\label{gauss}
 K=\frac{L'}{L} N' N - \frac{L''}{L} N^2 - (N')^2 + N N'' &=&
\gamma N^2\left(T^0_0 - T_{\varphi}^{\varphi}\right) + 2 NN' \frac{L'}{L} - 2 (N')^2 \nonumber \\
&=& \gamma N^2\left(\frac{2h^2 P^2}{L^2} + \frac{(P')^2}{L^2}\right) + 2 NN' \frac{L'}{L} - 2 (N')^2
\end{eqnarray}
where in the second last equality we have used the Einstein equations (\ref{N1}), (\ref{N2})
with $T_0^0$ the energy density and $T_{\varphi}^{\varphi}$ the pressure in
$\varphi$-direction.
Note that (\ref{gauss}) reduces to the relation found in \cite{gibbons} if we use
the assumptions made in that paper, i.e. $T^{\varphi}_{\varphi}=0$ and $N(\rho)\equiv 1$.

In the BPS limit we know that $N\equiv 1$ and the Gaussian curvature is obviously positive,
away from the BPS limit one has to use the numerical solution and compute the curvature.
We find that for most values of $\beta$ and $\gamma$
the Gaussian curvature is indeed positive and our result is in agreement with that of \cite{gibbons}. However, if the ratio between Higgs and gauge boson mass is sufficiently
large, we find that $K$ can become negative close to the string axis. Our results are
shown in Fig.\ref{gammabeta}, where we present $\gamma_{\rm cr}$ in dependence on
$\beta$. For $\gamma > \gamma_{\rm cr}$ the Gaussian curvature $K$ can become
negative for values of $\rho$ close to zero, while $K$ is strictly positive for
$\gamma < \gamma_{\rm cr}$. Though the theorem of \cite{gibbons} is not applicable here,
we nevertheless find that bound orbits do not exist.

\begin{figure}[h!]
\begin{center}
\resizebox{6in}{!}{\includegraphics{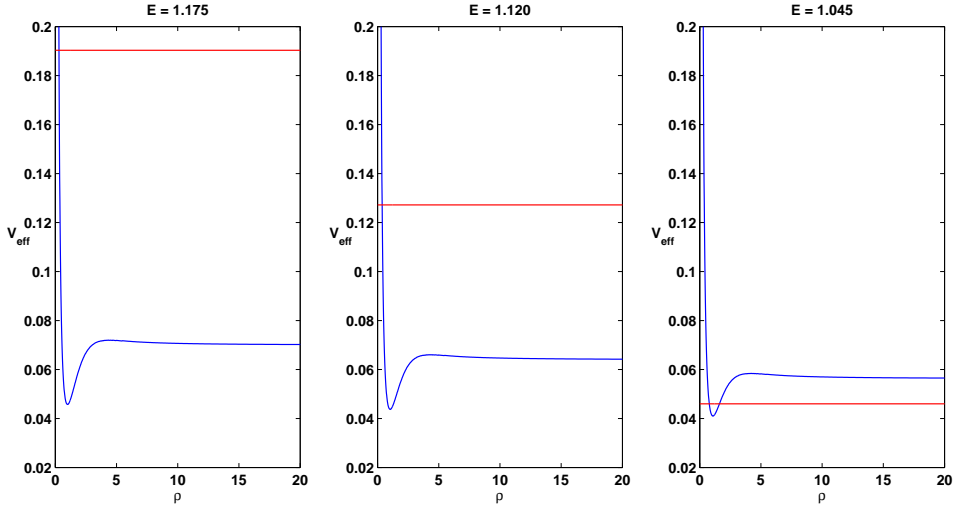}}
\end{center}
\caption{The effective potential for the motion of a massive test particle ($\varepsilon=1$) in the space--time of an Abelian--Higgs string with $\beta= 0.50$ and $\gamma =0.40$ for different
values of $E$. Here $p_z= 0.100$ and $L_z =0.18$. The red horizontal line represents the value of
$\mathcal{E}$.}\label{fignew2}
\end{figure}

\begin{figure}[h!]
\begin{center}
\resizebox{6in}{!}{\includegraphics{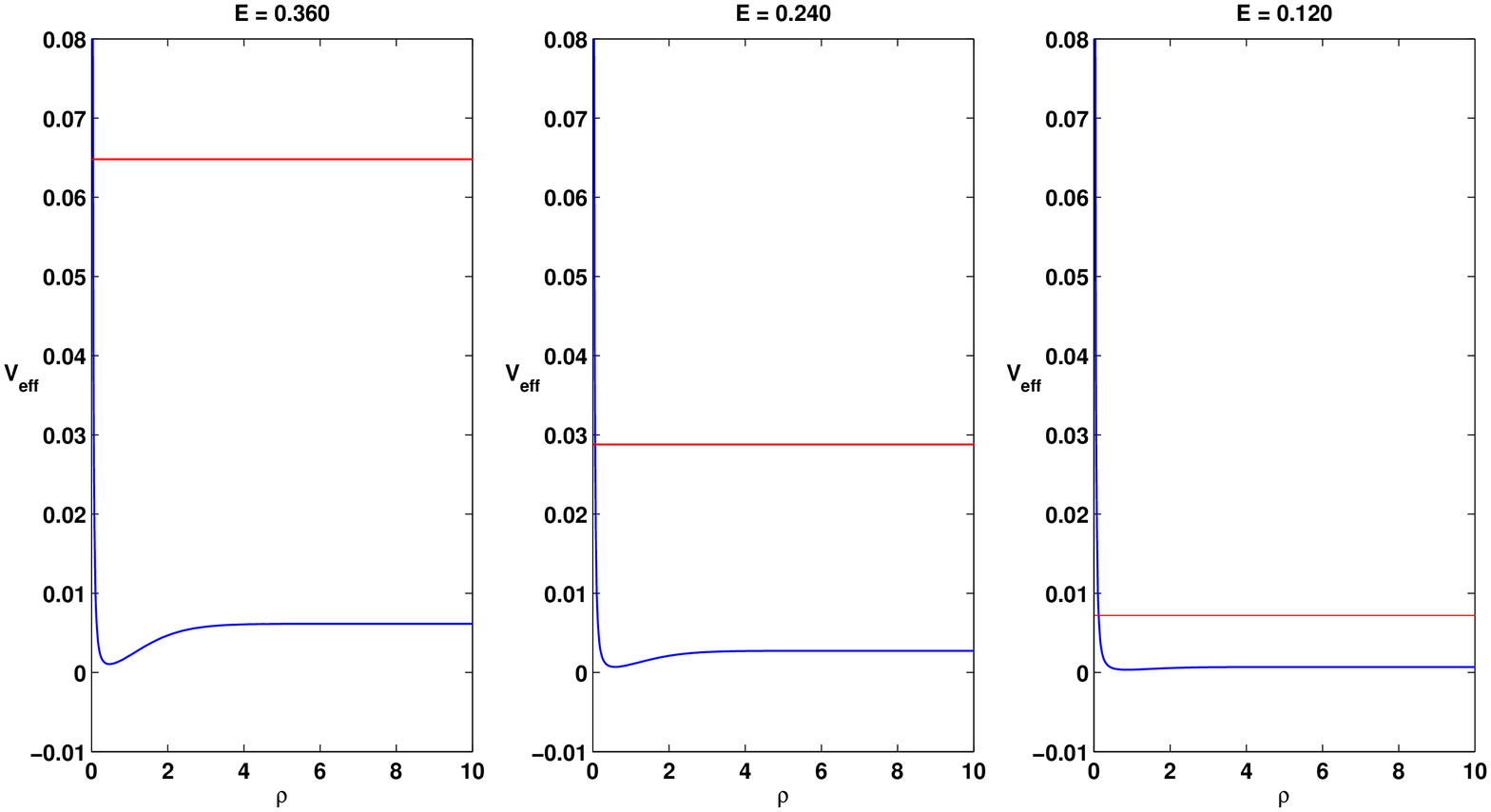}}
\end{center}
\caption{The effective potential for the motion of a massless test particle ($\varepsilon=0$) in the space--time of an Abelian--Higgs string with $\beta= 0.50$ and $\gamma =0.40$ for different
values of $E$. Here $p_z= 0.100$ and $L_z =0.015$. The red horizontal line represents the value of
$\mathcal{E}$.   }\label{fignew3}
\end{figure}

 \begin{figure}[h!]
\begin{center}
\resizebox{6in}{!}{\includegraphics{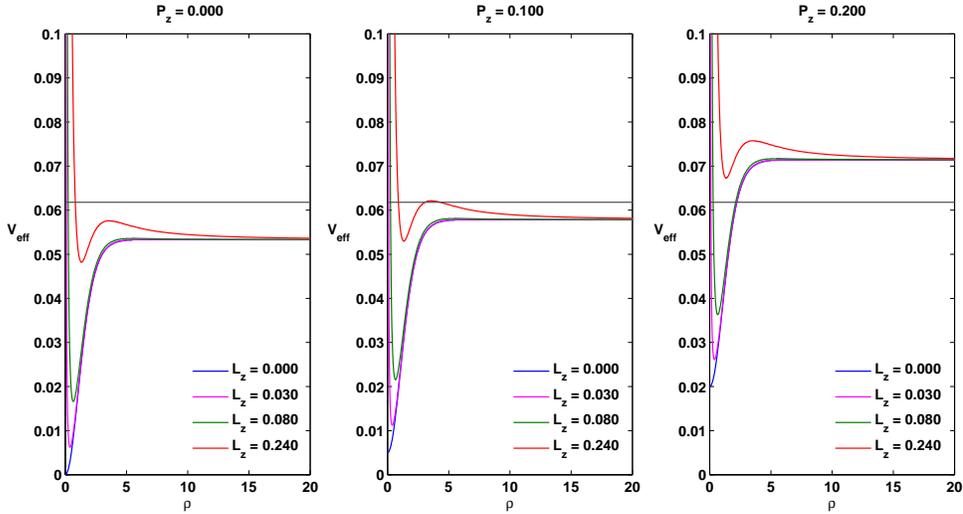}}
\end{center}
\caption{The effective potential for the motion of a massive test particle ($\varepsilon=1$) in the space--time of an Abelian--Higgs string with $\beta= 0.50$ and $\gamma= 0.40$ for different
values of $p_z$ and $L_z$. Here $E= 1.060$ and $L_z =0.015$. The black horizontal line represents the value of
$\mathcal{E}$. }\label{fignew4}
\end{figure}

In Fig.\ref{fignew4} we show how the effective potential changes with changing $p_z$ and $L_z$.
Increasing $p_z$ shifts the effective potential simply to higher values which
is apparent when looking at (\ref{potential}). Moreover, the lower the value of $L_z$ the lower
the value of the minimum of the potential. Moreover, the potential
barrier at $\rho=0$ disappears for $L_z=0$. 

\begin{figure}
\centering
\subfigure[$\mu$-$\nu$-plane for $\beta=0.5$ and $\gamma=0.4$]{\includegraphics[scale=0.45]{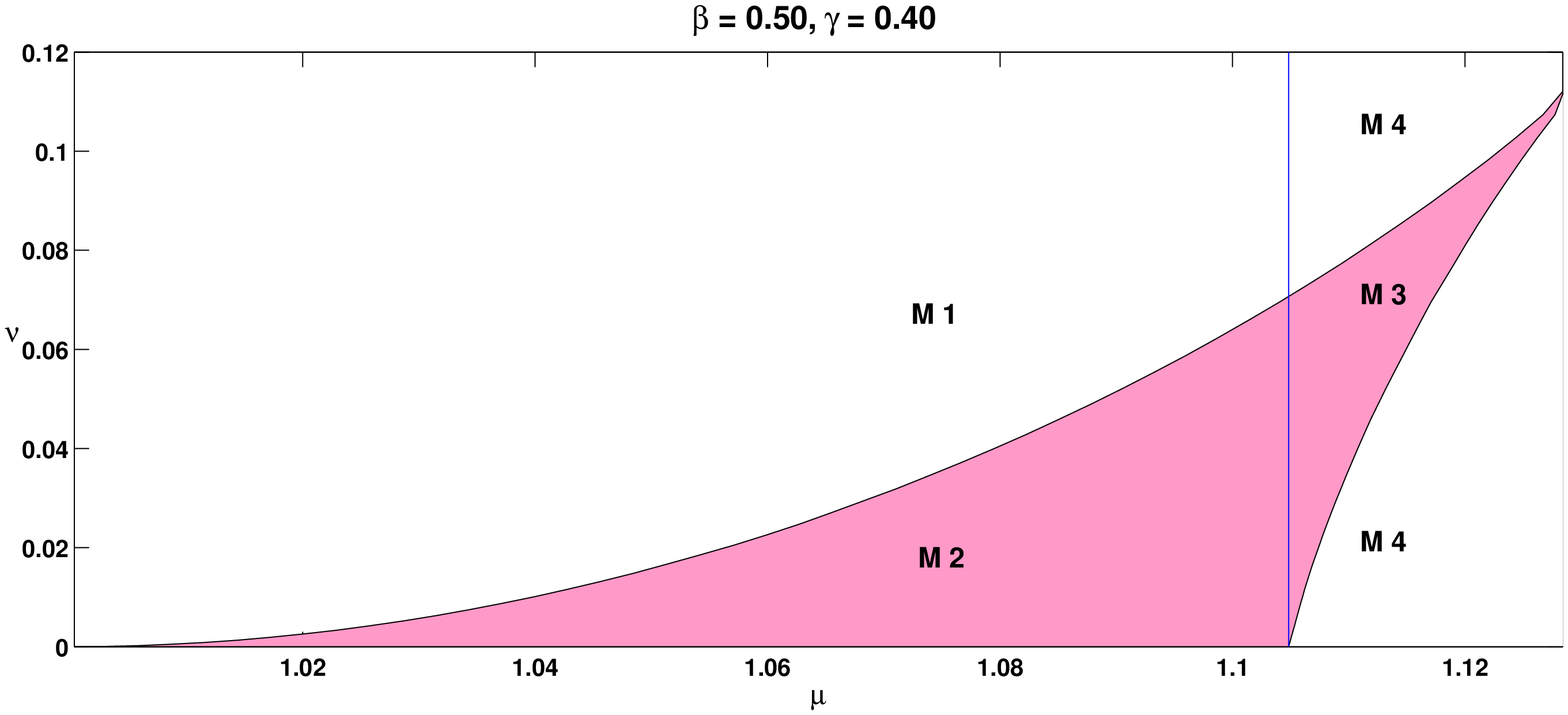}} \\
\subfigure[$\beta =1.9$, $\gamma=0.4$ ]{\includegraphics[scale=0.5]{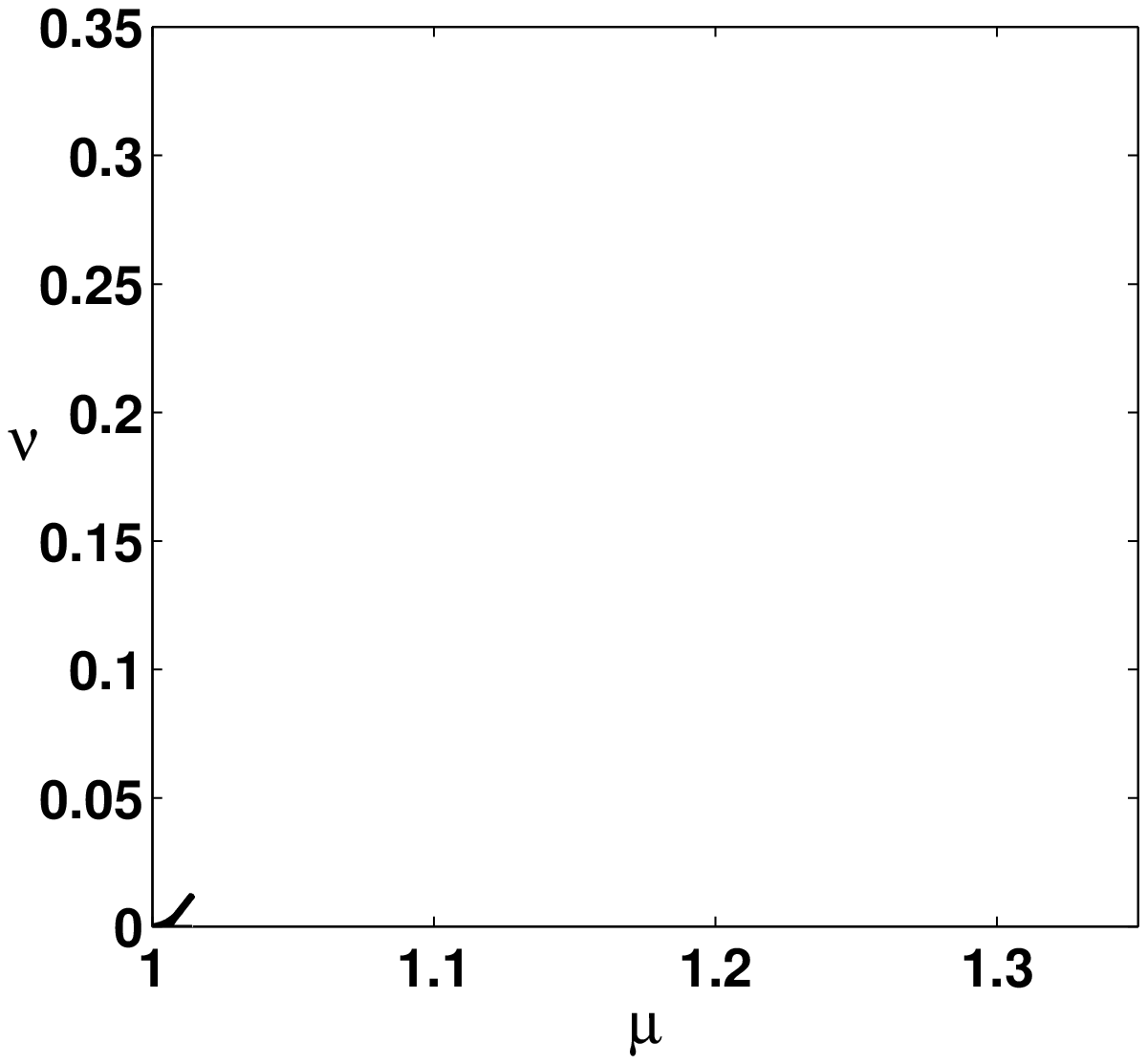}} 
\subfigure[$\beta =1.5$, $\gamma=0.4$ ]{\includegraphics[scale=0.5]{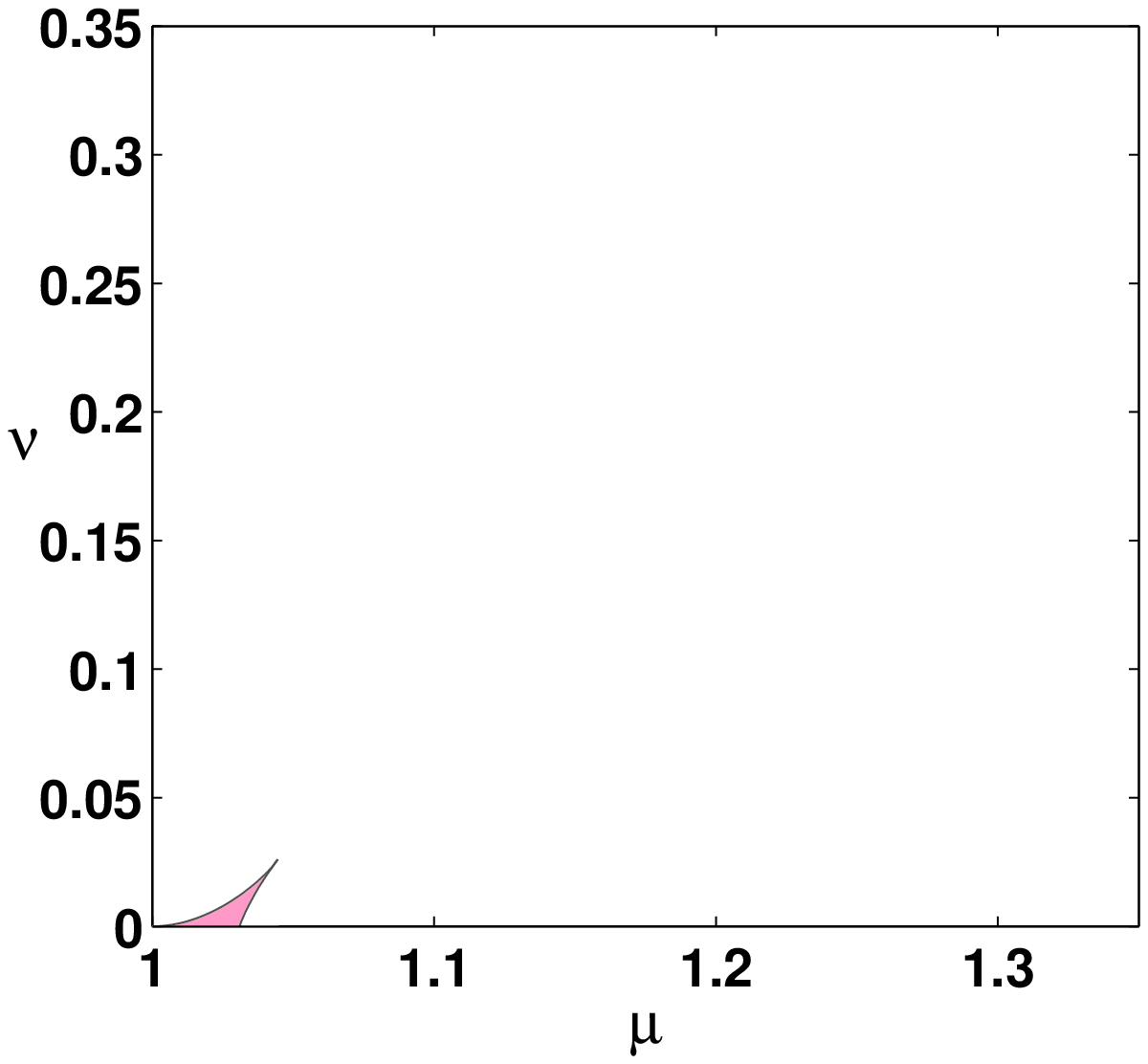}}
\\
\subfigure[$\beta =0.5$, $\gamma=0.2$ ]{\includegraphics[scale=0.5]{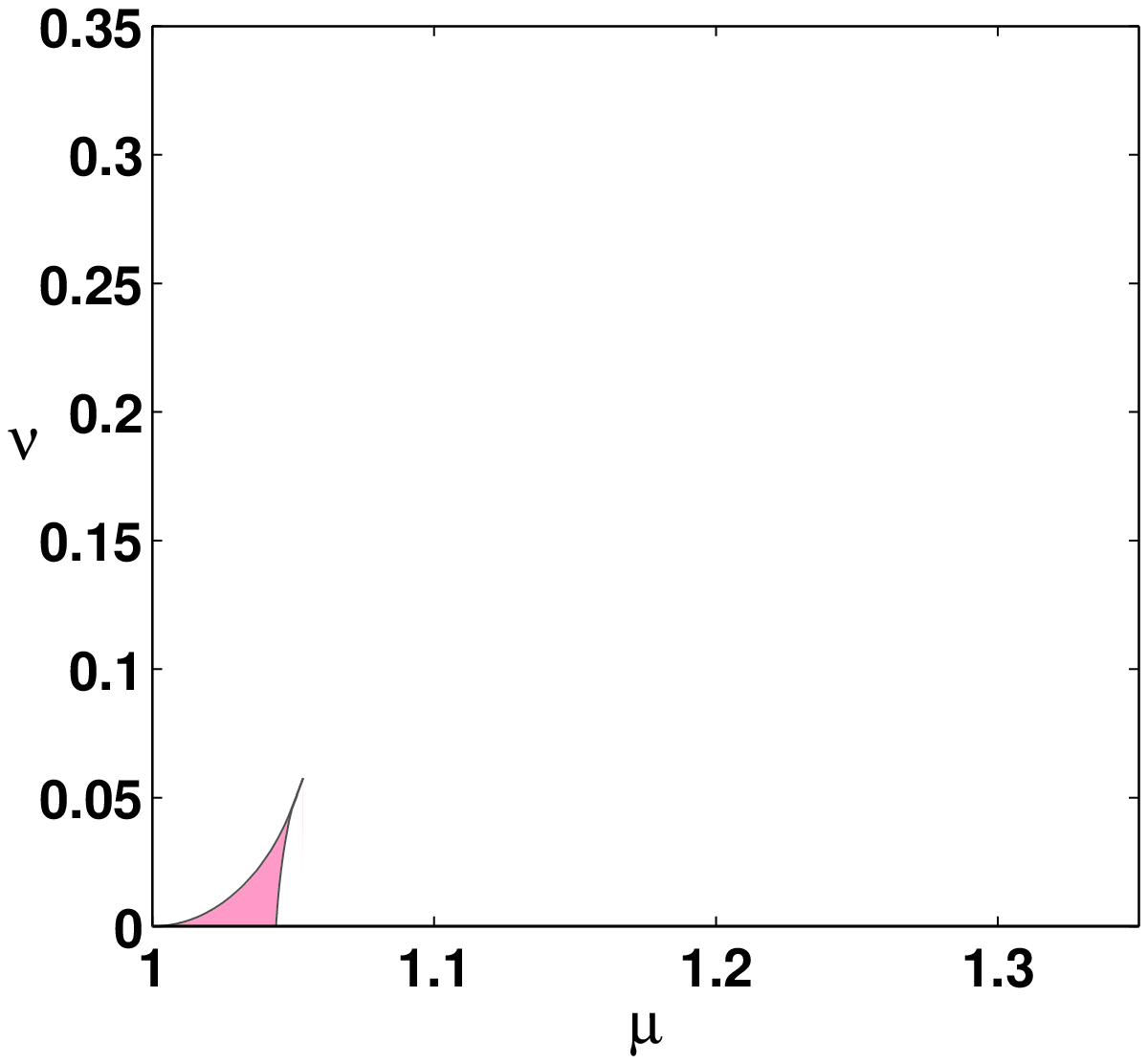}} 
\subfigure[$\beta =0.5$, $\gamma=0.6$ ]{\includegraphics[scale=0.5]{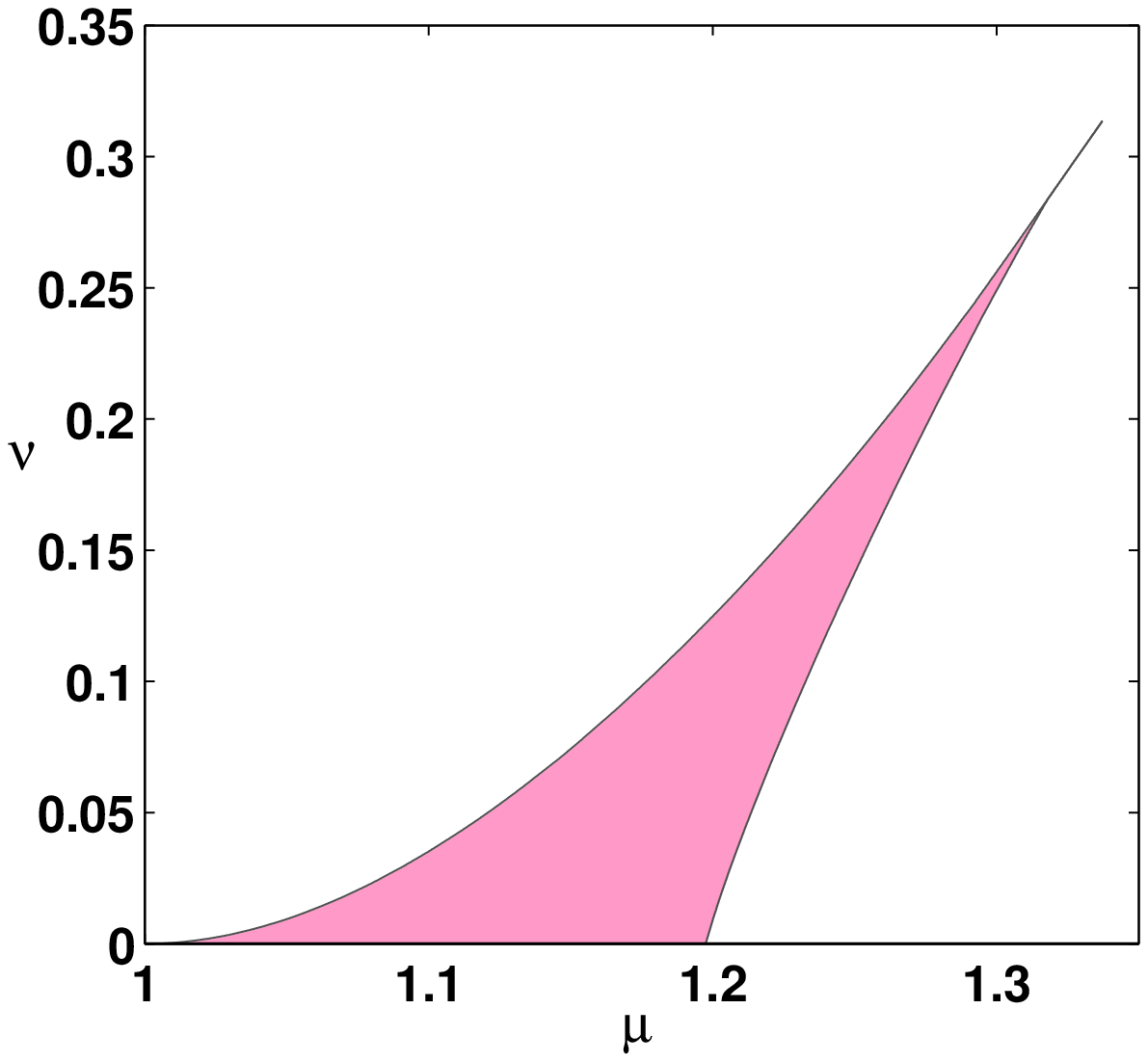}}
\caption{(a): The different regions in the $\mu$-$\nu$-plane for massive
test particles ($\varepsilon=1$) with $p_z=0$. Bound orbits are
possible in regions M2 and M3, while in M1 and M4 only escape orbits exist (see also Fig.\ref{domains} for more details). The blue vertical line indicates the value of the energy
at which the transition from region M3 to region M2 takes place.
(b)-(e): The change of the regions M1-M4 when changing the gravitational coupling $\gamma$ and the
Higgs to gauge boson mass ratio $\beta$, respectively. Note that for $\beta\ge 2$ no bound orbits
exist. }
\label{munu}
\end{figure}

\begin{figure}
\centering
\subfigure[Region M1]{\includegraphics[width=7cm]{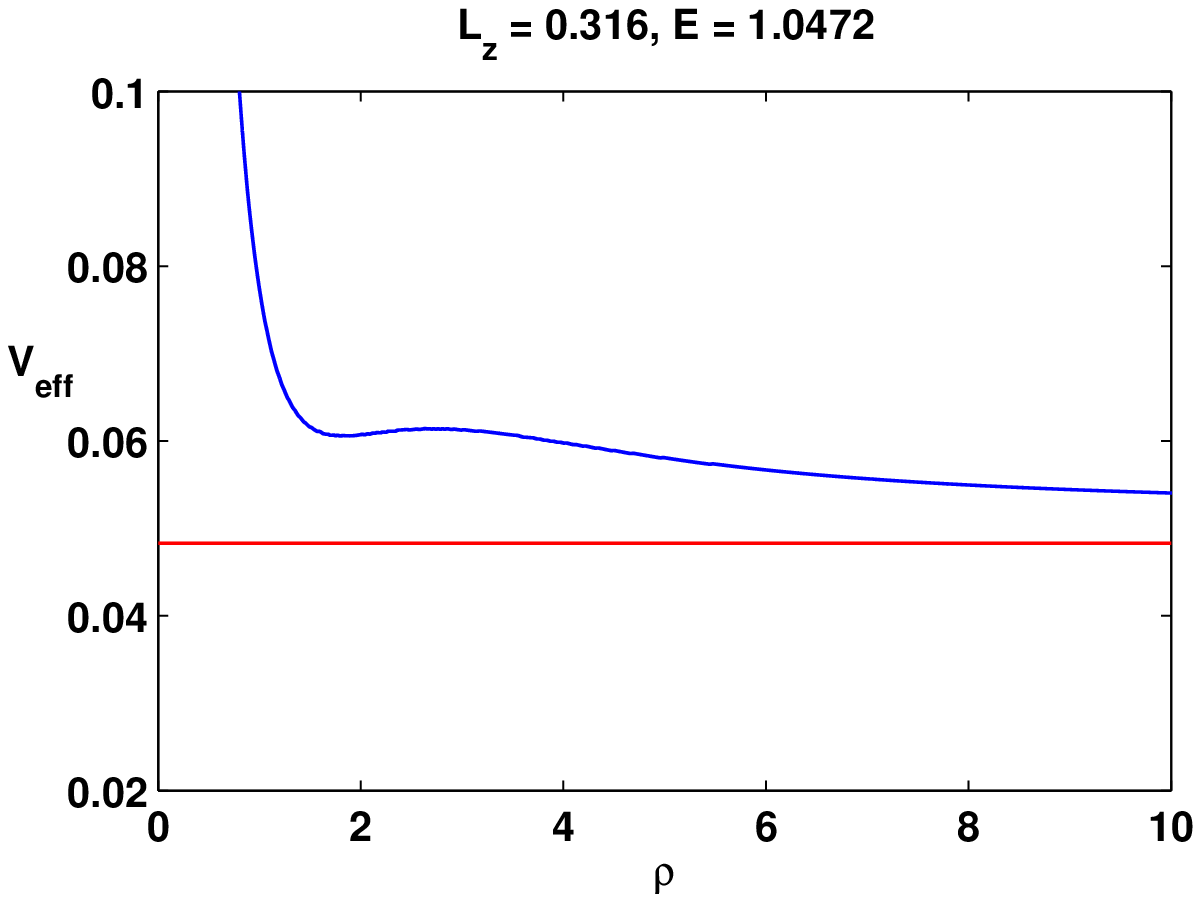}} 
\subfigure[Region M2]{\includegraphics[width=7cm]{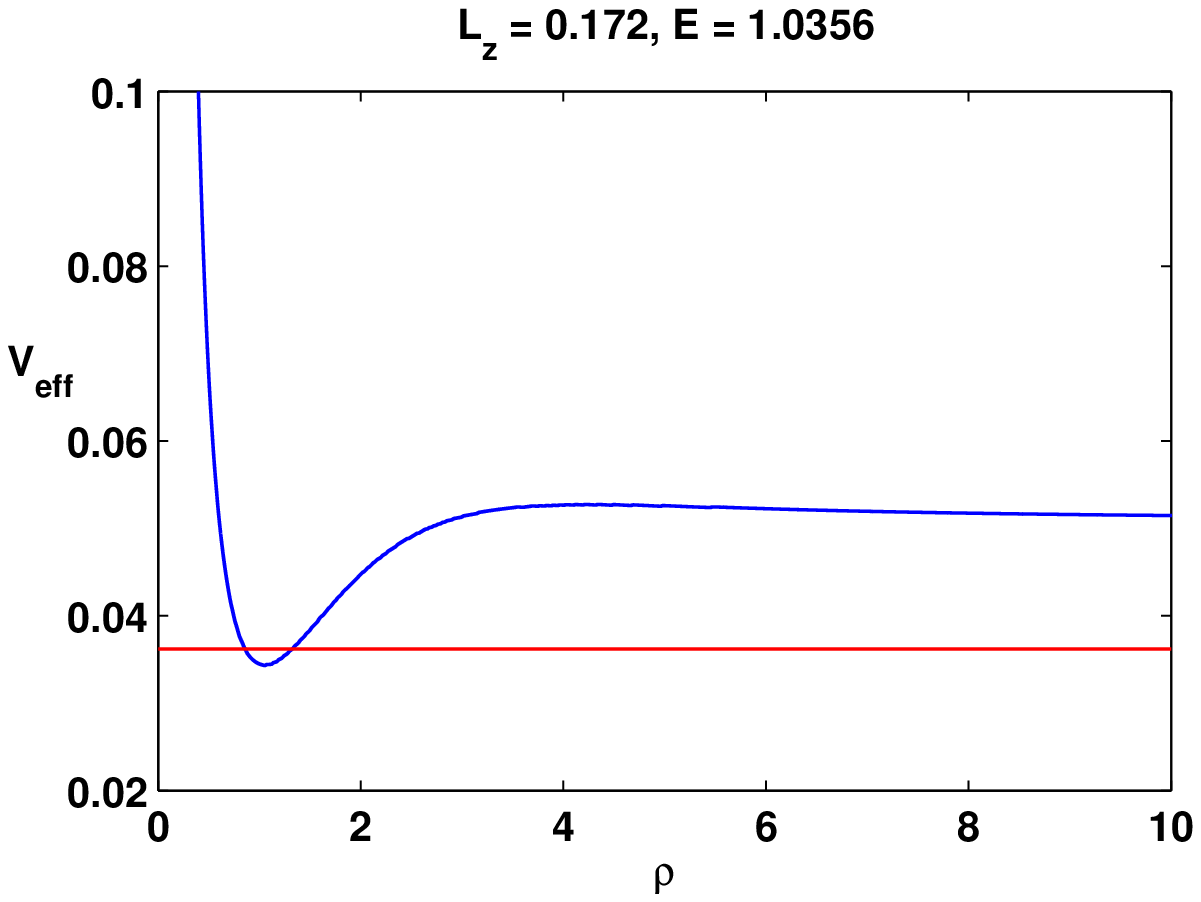}}
\\
\subfigure[Region M3]{\includegraphics[width=7cm]{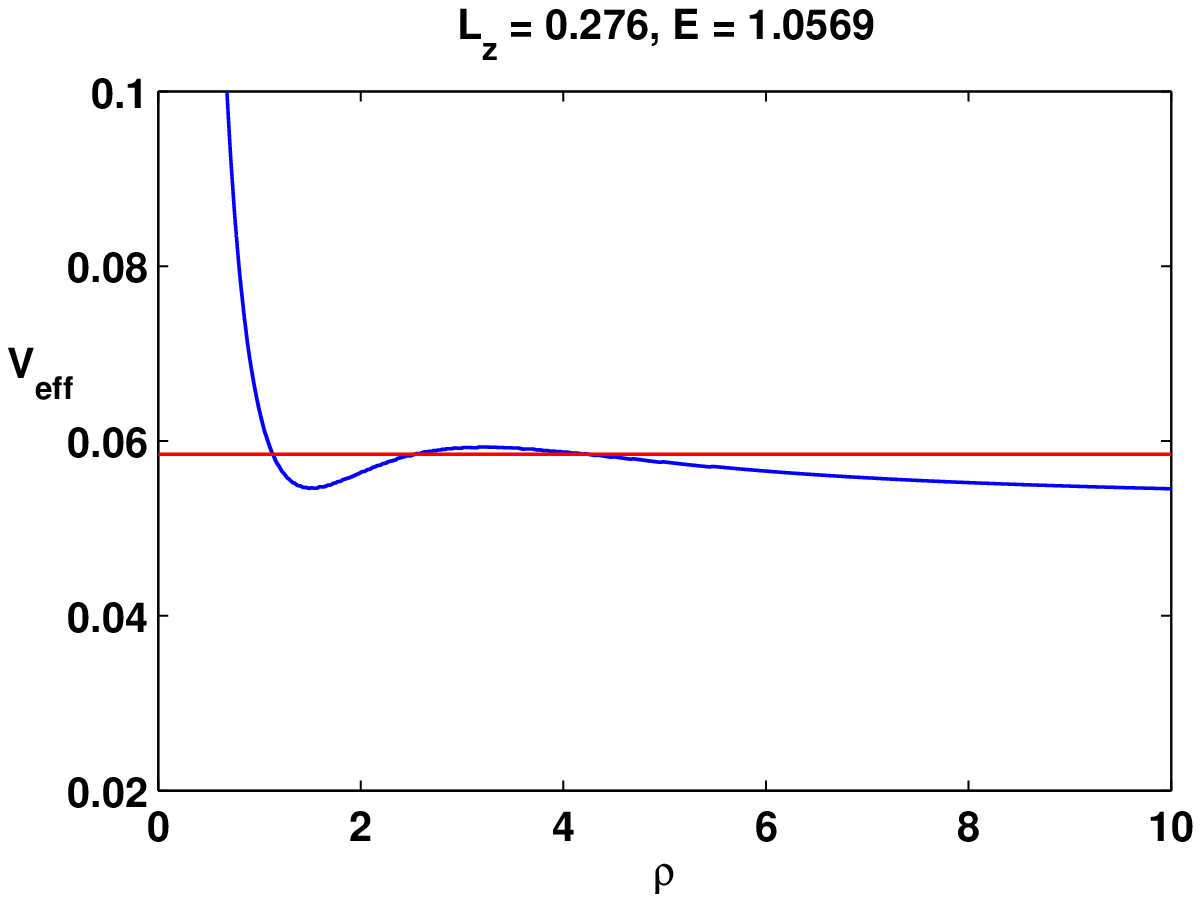}} 
\subfigure[Region M4]{\includegraphics[width=7cm]{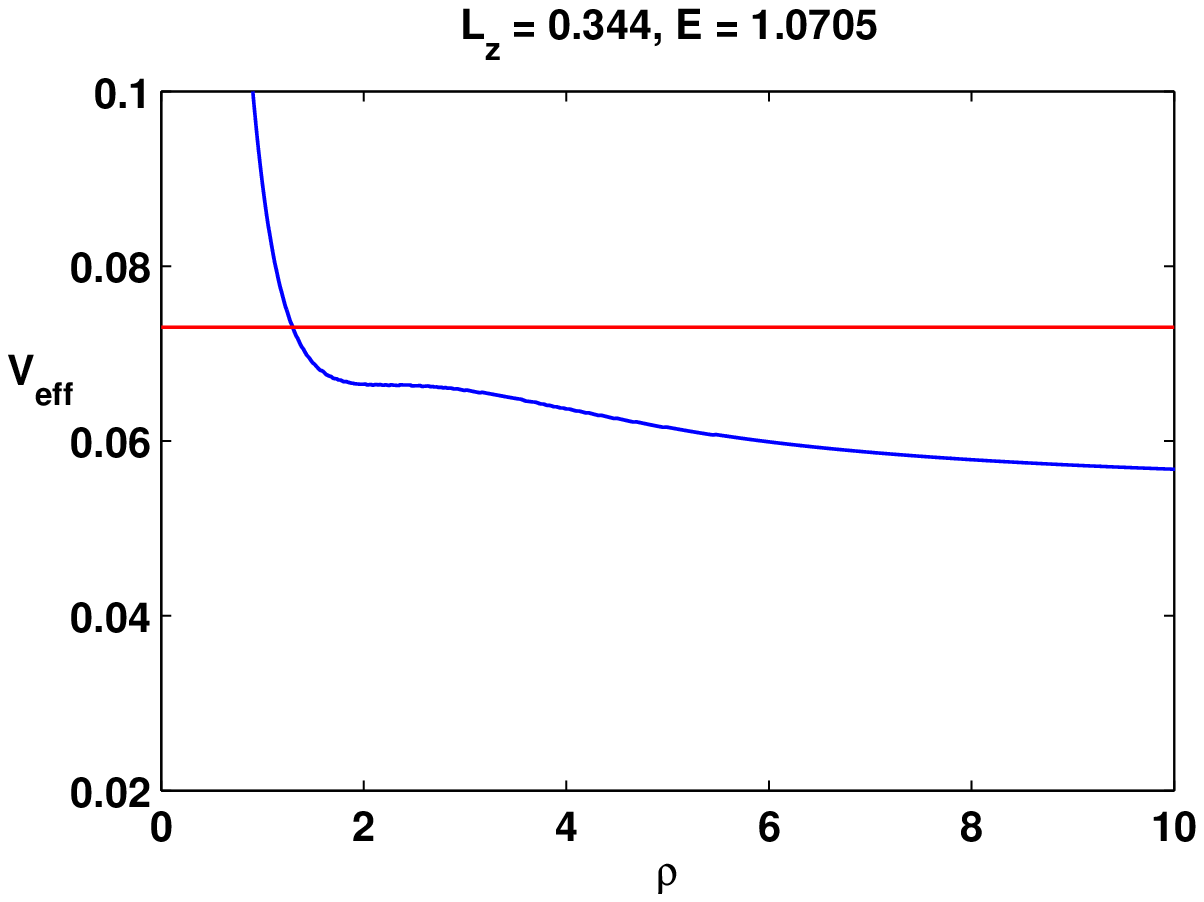}}
\caption{The effective potential $V_{\rm eff}(\rho)$ (blue) as well
as $(E^2-\varepsilon)/2$ (red) are given for the different regions M1-M4 of the $\mu$-$\nu$-plot (see
also Fig.\ref{munu}). Note that for particle motion to be possible we need to require $(E^2-\varepsilon)/2 > V_{\rm eff}(\rho)$. Hence in M2 there is a bound orbit, in M3 a bound and
an escape orbit, while in M4 there is an escape orbit. }
\label{domains}
\end{figure}

We have then investigated how the existence of escape and
bound orbits  depends on the choice of $E$, $L_z$ and $p_z$. Following the known
discussion of orbits in the Schwarzschild case \cite{hagi}, we
define $\nu:=L_z^2$ and $\mu:=E^2$.
In Fig.\ref{munu}(a) we show the $\mu$--$\nu$--plane of a massive
test particle for $\beta=0.5$, $\gamma=0.4$ and $p_z=0$ and indicate for
which values of $\mu$ and $\nu$ bound orbits exist. This is only possible
in regions $M2$ and $M3$. The corresponding effective
potential $V_{\rm eff}$ for regions M1-M4 is shown in Fig.\ref{domains}.
In region M1 the effective potential $V_{\rm eff}$ is always 
larger than $\cal{E}$, so no particle motion is possible.
In region M2 the potential has a local minimum and $\cal{E}$ intersects
$V_{\rm eff}$ twice. The intersection points correspond to the
minimal and maximal radius of a bound orbit.
In region M3  the potential has a local minimum and a local maximum
and $\cal{E}$ intersects $V_{\rm eff}$ three times.
The two intersection points at smaller $\rho$ correspond to the
minimal and maximal radius of a bound orbit, while the intersection
point at larger $\rho$ corresponds to the minimal
radius of an escape orbit. Finally, in region M4 $\cal{E}$ intersects $V_{\rm eff}$ once. This intersection point corresponds to the minimal
radius of an escape orbit.
As already mentioned the form of the effective potential will
change with $\gamma$ and $\beta$. This leads then to a change of
the $\mu$--$\nu$--plane which is indicated in Fig.\ref{munu}(b)-(e), where
we demonstrate how the regions M1-M4 change when changing either $\gamma$ or $\beta$. For fixed $\gamma$ the regions M2 and M3 in which bound orbits
exist get smaller when increasing $\beta$ and in fact -- as stated above --
disappear completely for $\beta \ge 2$. This is seen in Fig.\ref{munu}(b)-(c). Moreover, for fixed $\beta < 2$ the regions
M2 and M3 get bigger when increasing the gravitational coupling
$\gamma$. This is shown in Fig.\ref{munu}(d)-(e).

Note that our results for $p_z\neq 0$ are qualitatively similar, this is why we don't discuss them in detail
here. The main difference between $p_z=0$ and $p_z\neq 0$ is that the particles
move on 3--dimensional orbits in the latter case, while their motion is restricted
to the $x$--$y$--plane in the former.

\subsection{Examples of geodesics}
\subsubsection{Massive particles $\varepsilon=1$}

In Fig.\ref{geo1}-\ref{geo4} we show how bound and escape orbits of massive test particles $(\varepsilon=1)$ change
when changing $\gamma$ and $\beta$, respectively. In order to understand how the test particle
moves we also indicate the radius of the scalar core (red) and of the magnetic flux tube core (blue).
Since we measure the distance $\rho$ in units of $M_{\rm W}/\sqrt{2}$ the scalar core
radius is given by $\rho_{\rm H}\sim 1/\sqrt{\beta}$ and the magnetic flux tube core is given by $\rho_{\rm W}\sim 1/\sqrt{2}$.

From Fig.\ref{geo1} it is evident that the maximal radius of a bound orbit decreases
strongly with increasing $\gamma$. While it is much larger than $\rho_{\rm H}$ for
$\gamma=0.36$, it becomes comparable to $\rho_{\rm H}$ for $\gamma=0.42$ and even
smaller than $\rho_{\rm H}$ for $\gamma=0.48$. This is related to the increased
curvature of space--time for increasing $\gamma$. Moreover, the minimal
radius of the escape orbit is always inside the magnetic flux tube core. Hence, the particle
does not only move in the exterior (vacuum) region of the cosmic string, but can
enter into the scalar and flux tube cores. 

The change of a bound orbit with $\beta$ is shown in Fig.\ref{geo2}. The maximal
radius of the bound orbit increases with increasing $\beta$, while the radius of
the scalar core $\rho_{\rm H}$ decreases at the same time. This leads to the observation
that for small $\beta$ the particle moves basically only inside the scalar core,
while for increasing $\beta$ it moves further and further away from the cosmic string.

\begin{figure}[htp]
\centering

\subfigure[$\gamma$ = 0.36]{
\includegraphics[scale=0.5]{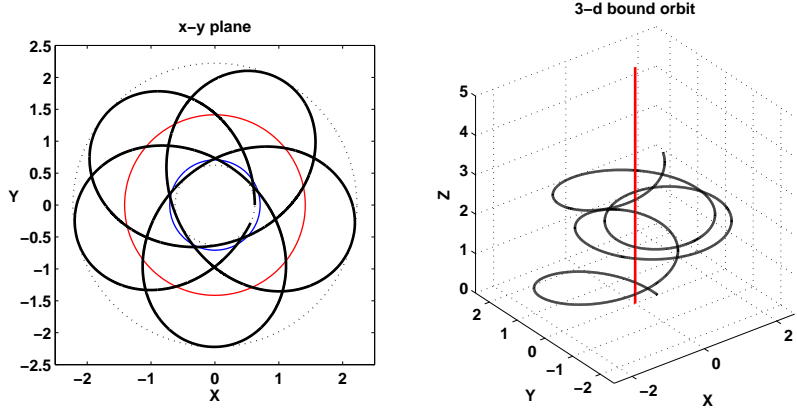}

}

\subfigure[$\gamma$ = 0.42]{
\includegraphics[scale=0.5]{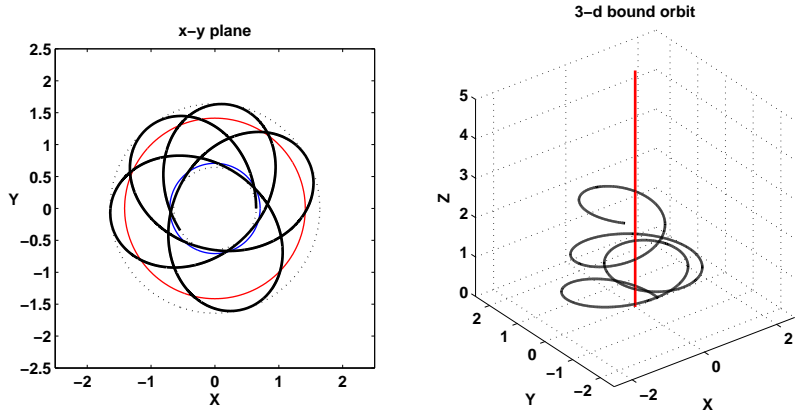}

}
\subfigure[$\gamma$ = 0.48]{
\includegraphics[scale=0.5]{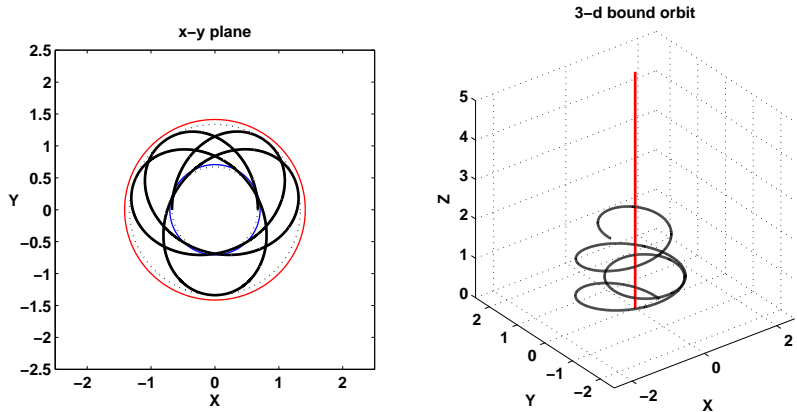}

}

\caption[Optional caption for list of figures]{The change of the bound orbit of
a massive test particle ($\varepsilon=1$) with $E=1.083$, $L_z^2=0.025$ and $p_z=0.02$ 
due to the change of the gravitational coupling $\gamma$. Here, we have chosen
$\beta=0.5$. The red and blue circle indicates the scalar field
core and the magnetic flux tube core of the Abelian--Higgs string, respectively. The dotted
lines indicate the minimal and maximal $\rho$ of the bound orbit. In the 3--dimensional
plots the red line indicates the string axis.}\label{geo1}
\end{figure}

\begin{figure}[htp]
\centering

\subfigure[$\beta$ = 0.40]{
\includegraphics[scale=0.5]{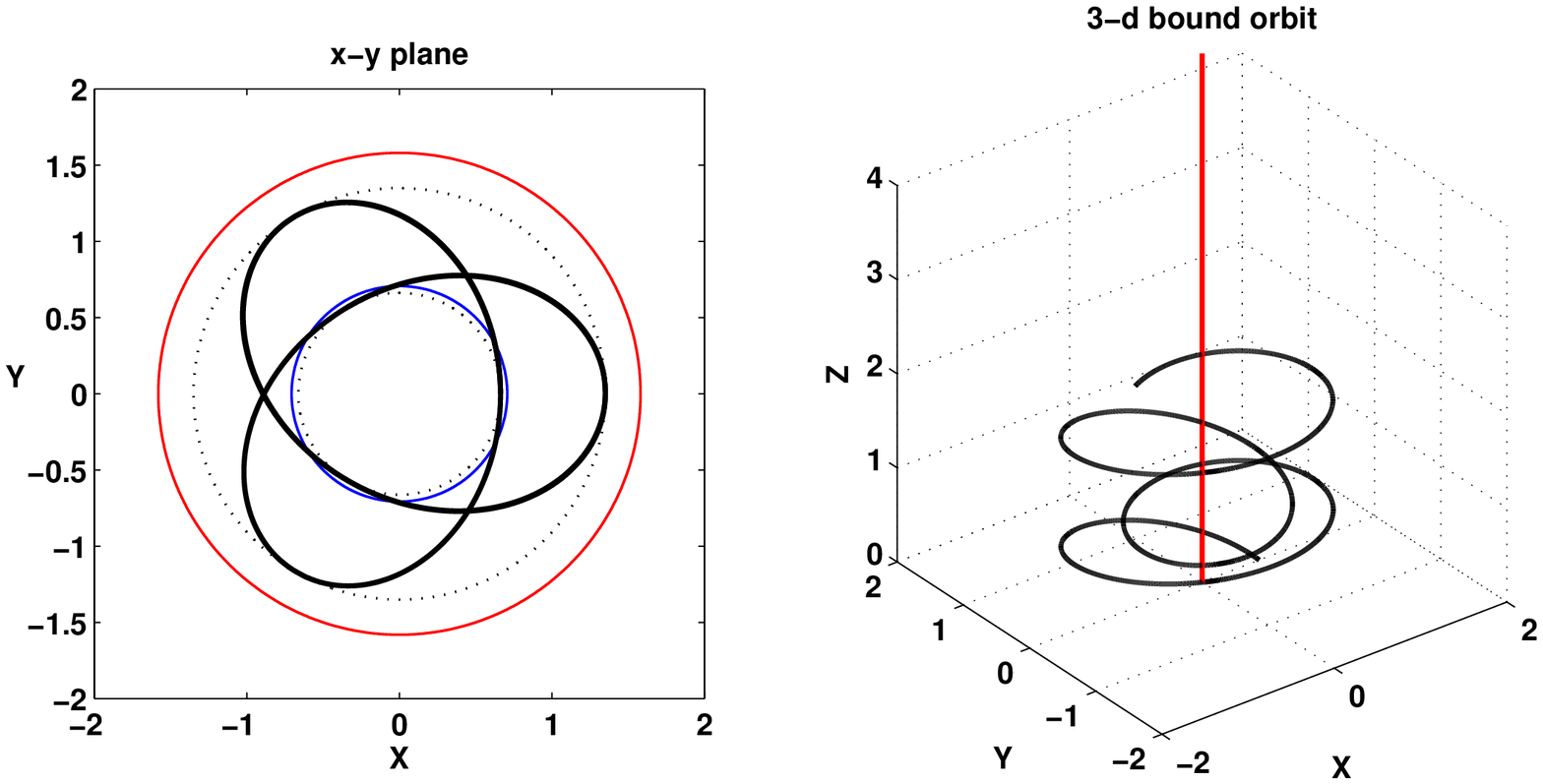}
}

\subfigure[$\beta$ = 0.50]{
\includegraphics[scale=0.5]{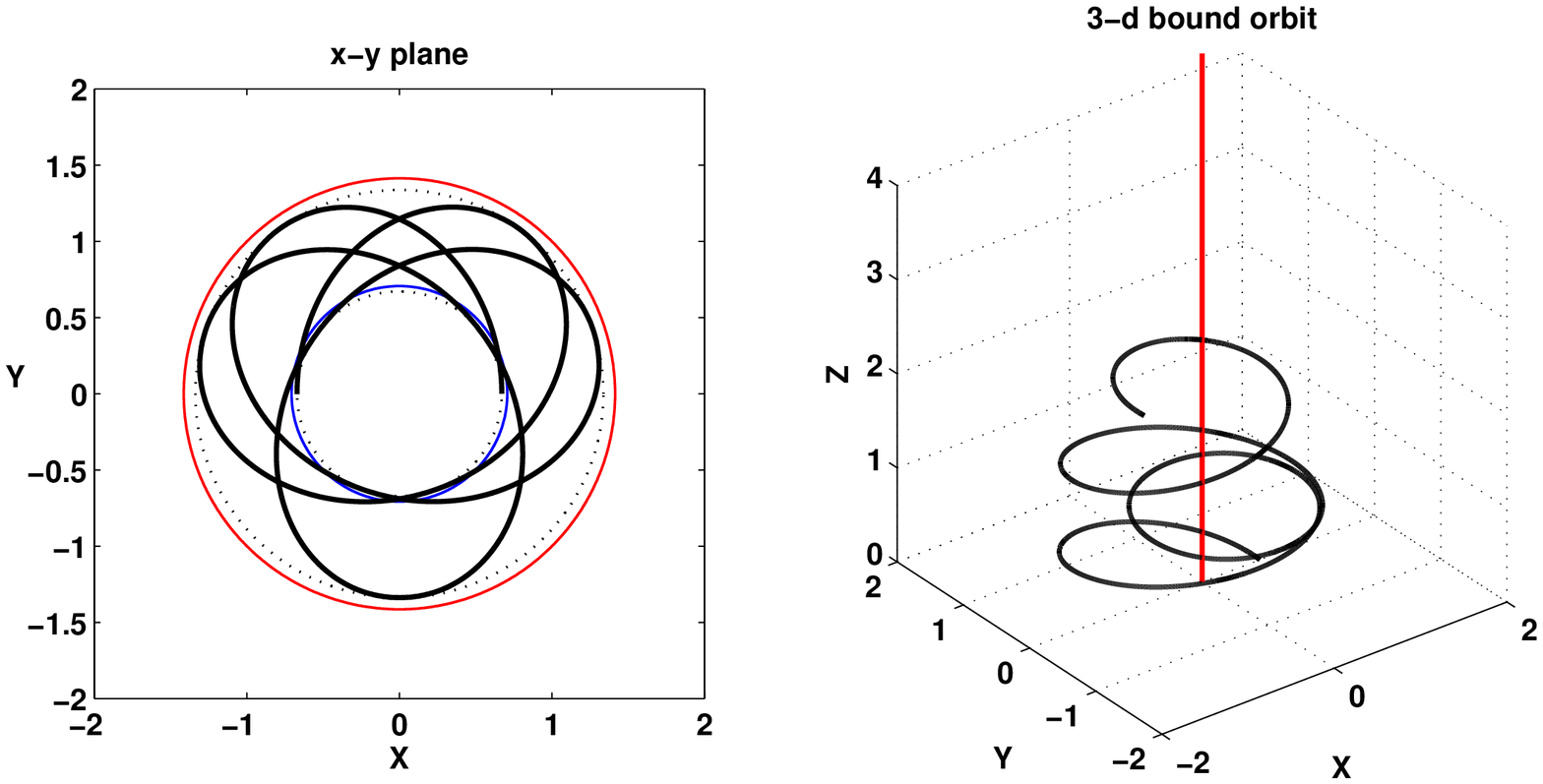}
}
\subfigure[$\beta$ = 0.60]{
\includegraphics[scale=0.5]{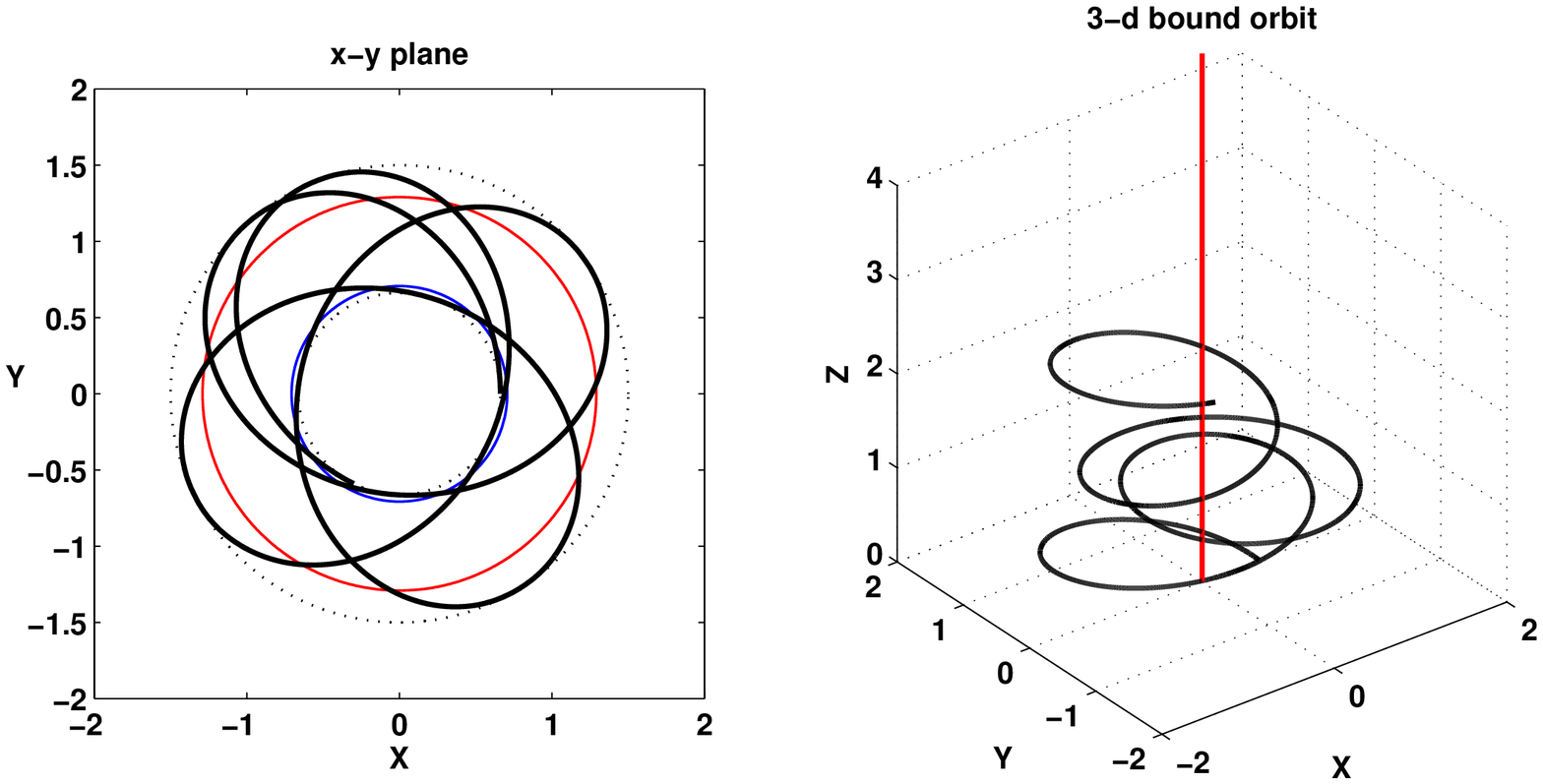}
}

\caption[Optional caption for list of figures]{The change of the bound orbit of
a massive test particle ($\varepsilon=1$) with $E=1.083$, $L_z^2=0.025$ and $p_z=0.0205$ 
due to the change of the ratio between Higgs and gauge boson mass $\beta$. Here, we have chosen
$\gamma=0.48$. The red and blue circle indicates the scalar field
core and the magnetic flux tube core of the Abelian--Higgs string, respectively. The dotted
lines indicate the minimal and maximal $\rho$ of the bound orbit. In the 3--dimensional
plots the red line indicates the string axis.}\label{geo2}
\end{figure}

In Fig.\ref{geo3} we show how an escape orbit changes when changing the gravitational
coupling $\gamma$.  For $\gamma=0.15$ and $\gamma=0.3$ the particle arrives
from infinity, gets deflected by the cosmic string and moves again to infinity.
The larger $\gamma$ the bigger is the change in the direction of the motion of the particle.
This is not suprising since the deficit angle $\Delta$ increases with increasing $\gamma$.
For $\gamma=0.45$ we observe a new phenomenon. The particle arrives from infinity, encircles the cosmic string and then moves again to infinity. This is new as compared
to a space--time of an infinitely thin string. However, this type of encirclement has
already been observed in the space--time of a Schwarzschild black hole and Kerr black hole, respectively pierced
by an infinitely thin cosmic string for sufficiently large deficit angle \cite{hhls1,hhls2}.
Comparing these results with those given in Fig.\ref{geo4} we observe that
the encirclement disappears when increasing $\beta$. While it is still
present for $\beta=0.5$, it has disappeared for $\beta=2$ and $\beta=5.3$.

\begin{figure}[htp]
\centering

\subfigure[$\gamma= 0.15$]{
\includegraphics[scale=0.50]{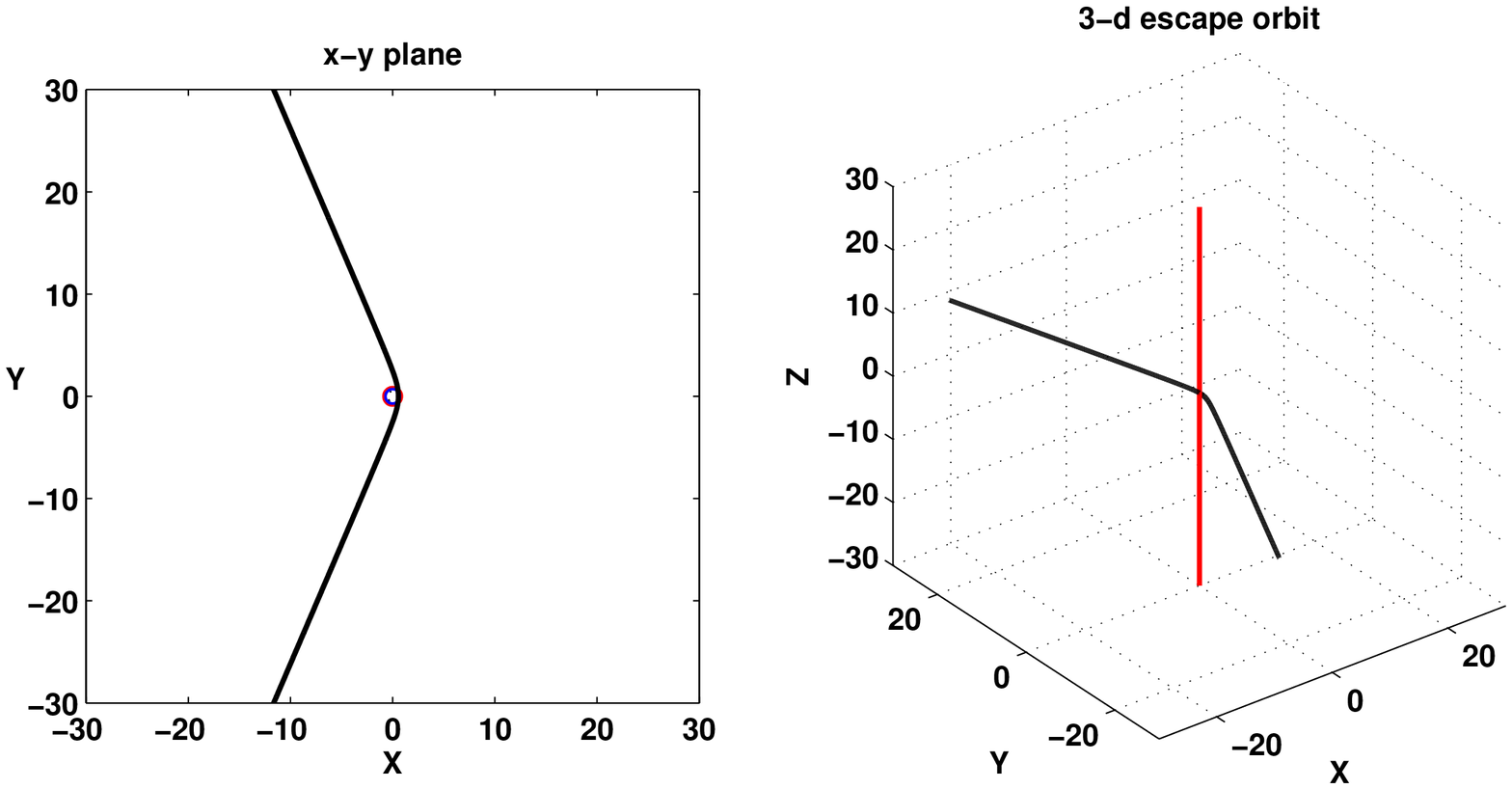}

}

\subfigure[$\gamma = 0.30$]{
\includegraphics[scale=0.50]{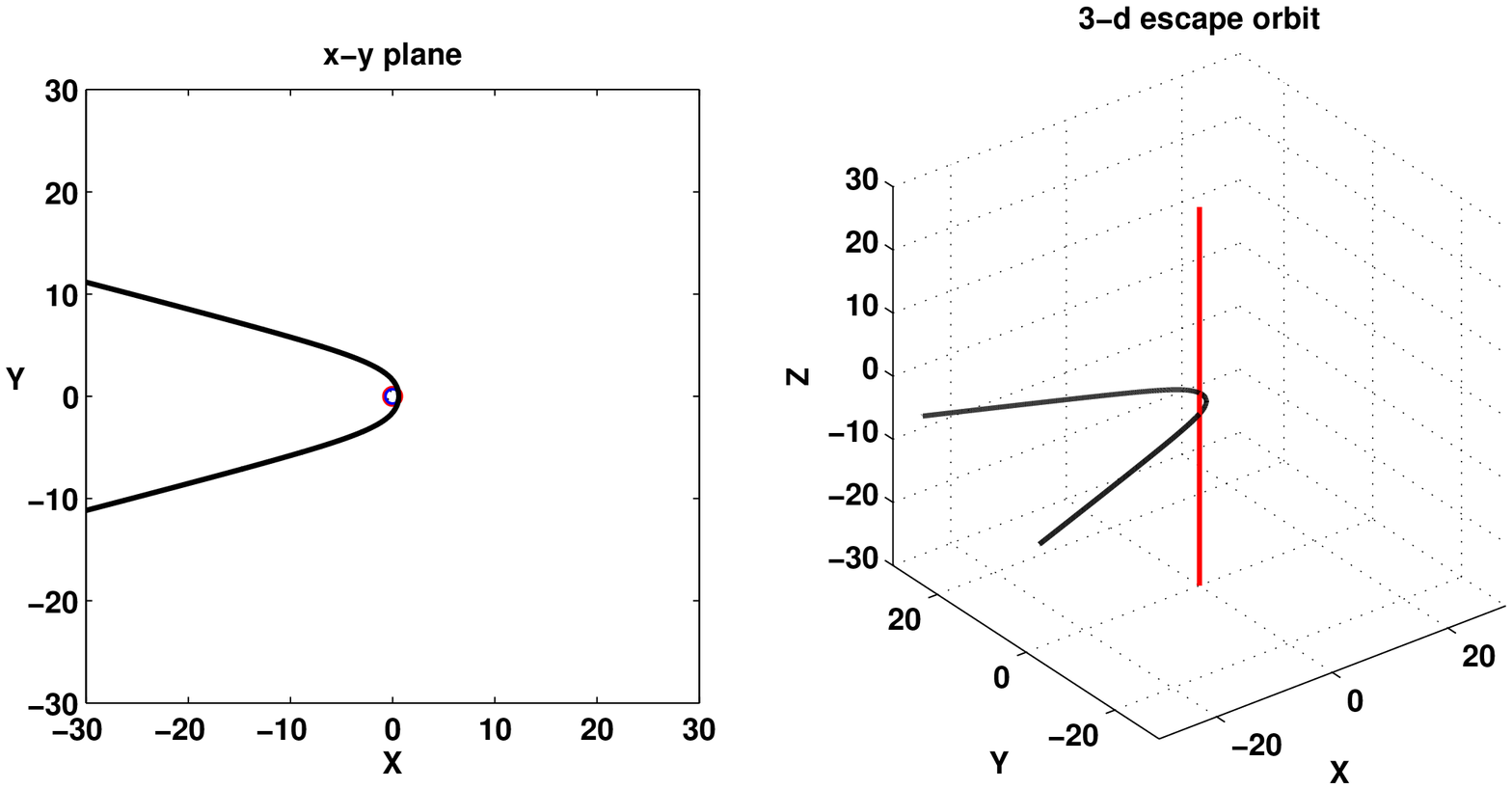}

}
\subfigure[$\gamma = 0.45$]{
\includegraphics[scale=0.50]{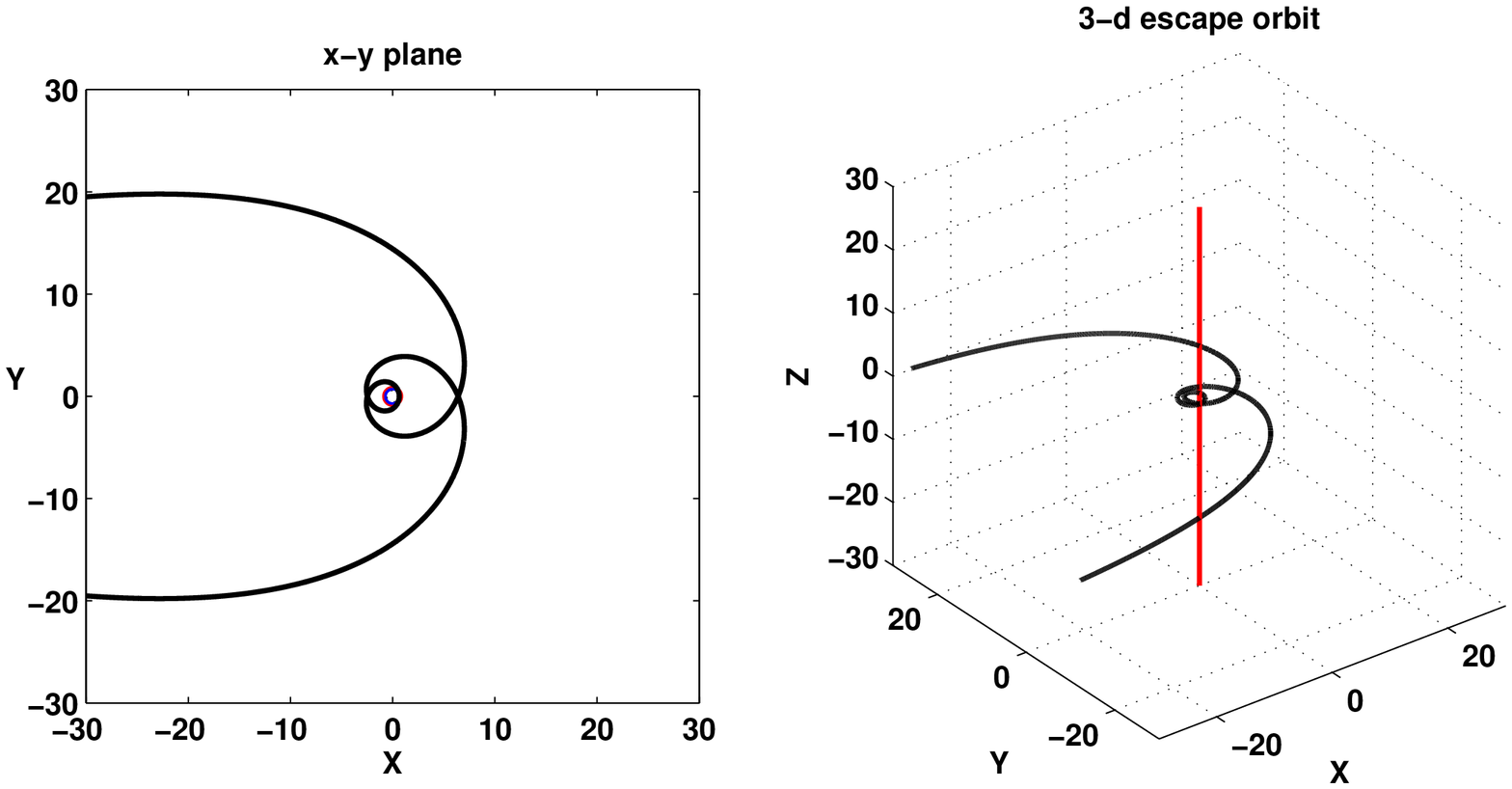}

}
\caption[Optional caption for list of figures]{The change of the escape orbit of
a massive test particle ($\varepsilon=1$) with $E=1.034$, $L_z^2=0.02$ and $p_z=0.002$ 
due to the change of the gravitational coupling $\gamma$. Here, we have chosen
$\beta=1.25$. The red and blue circle indicates the scalar field
core and the magnetic flux tube core of the Abelian--Higgs string, respectively. In the 3--dimensional
plots the red line indicates the string axis.}\label{geo3}
\end{figure}

\begin{figure}[htp]
\centering

\subfigure[$\beta = 0.50$]{
\includegraphics[scale=0.5]{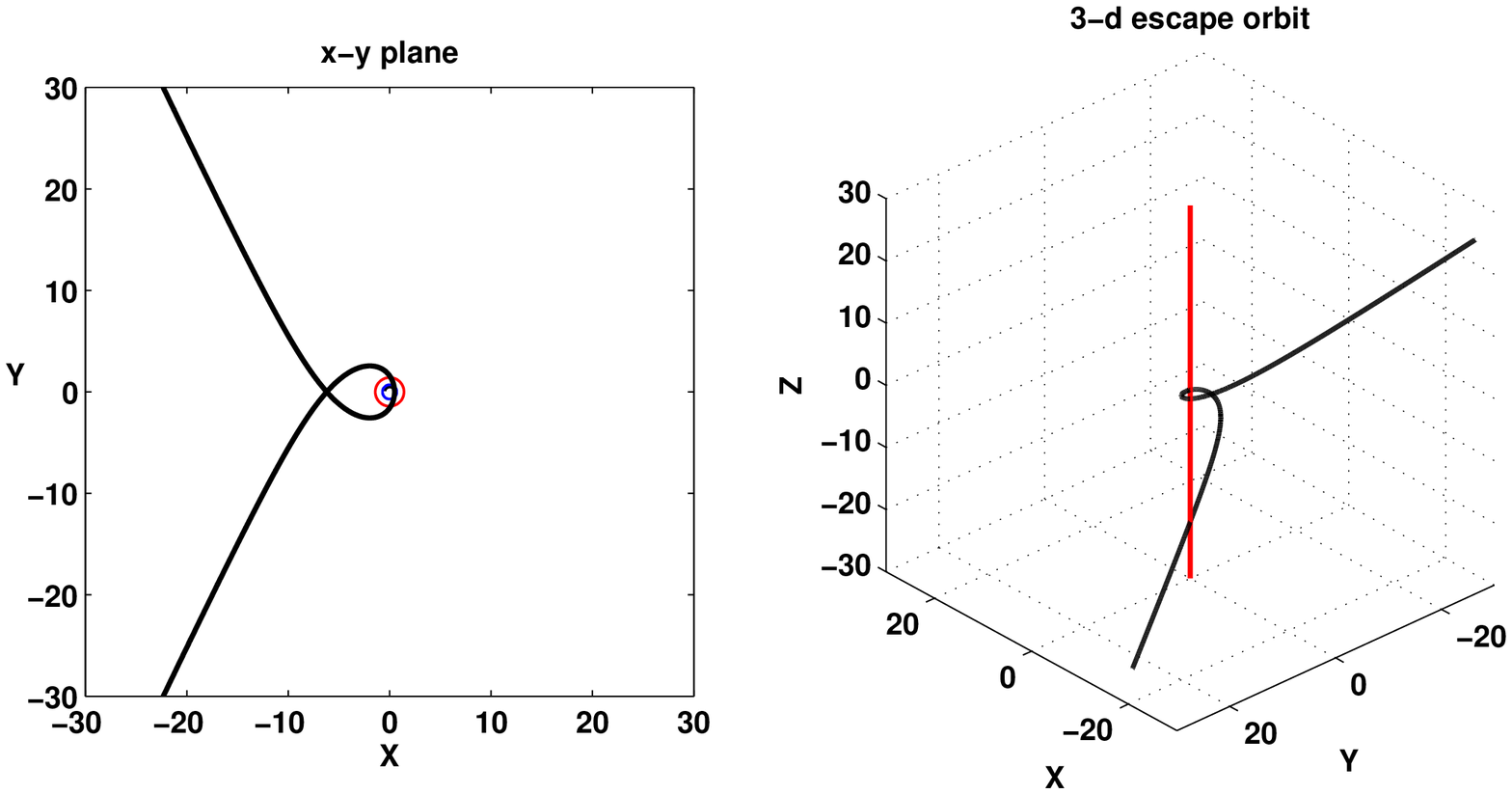}

}

\subfigure[$\beta = 2.00$]{
\includegraphics[scale=0.5]{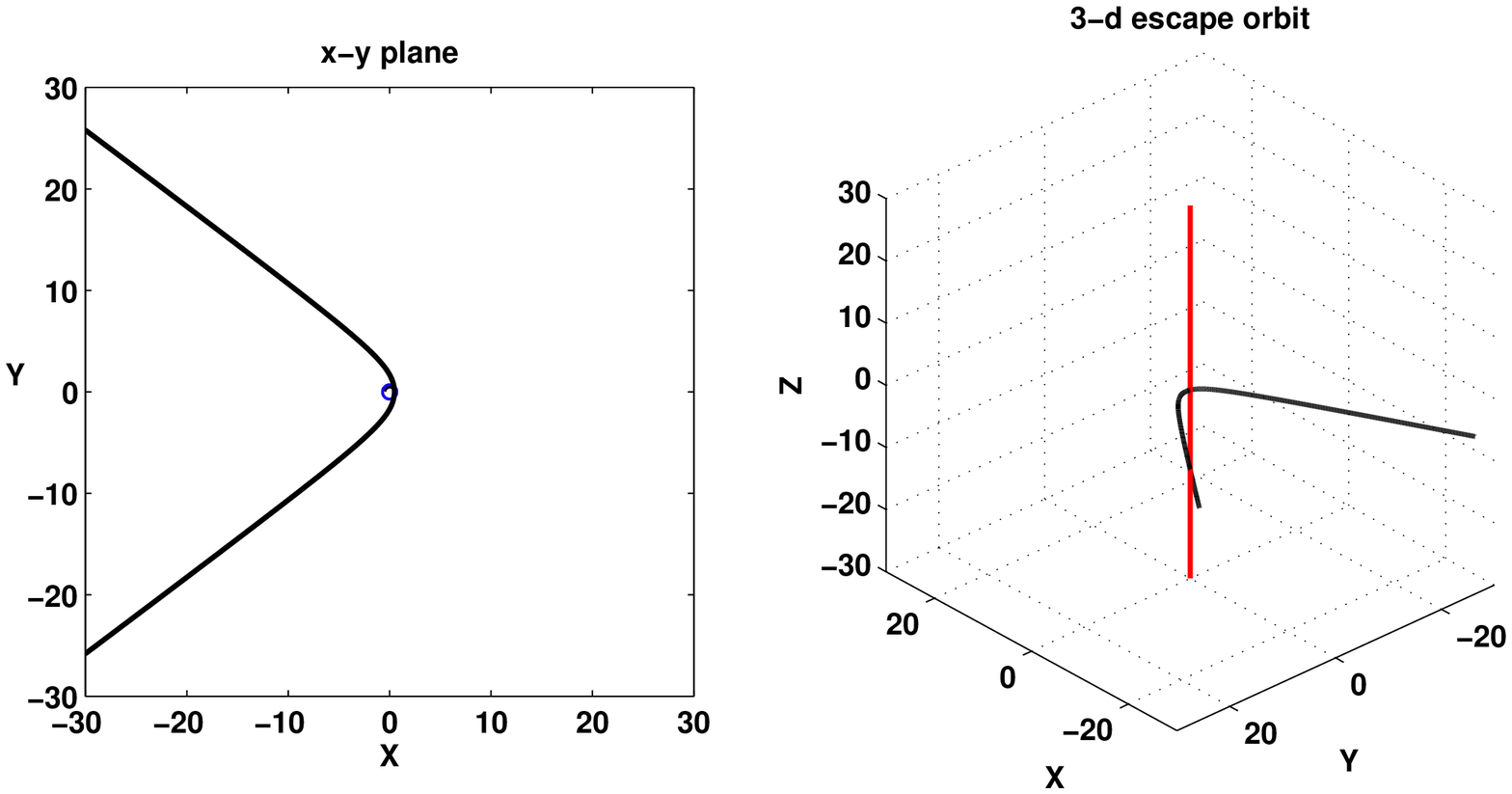}

}
\subfigure[$\beta = 5.30$]{
\includegraphics[scale=0.5]{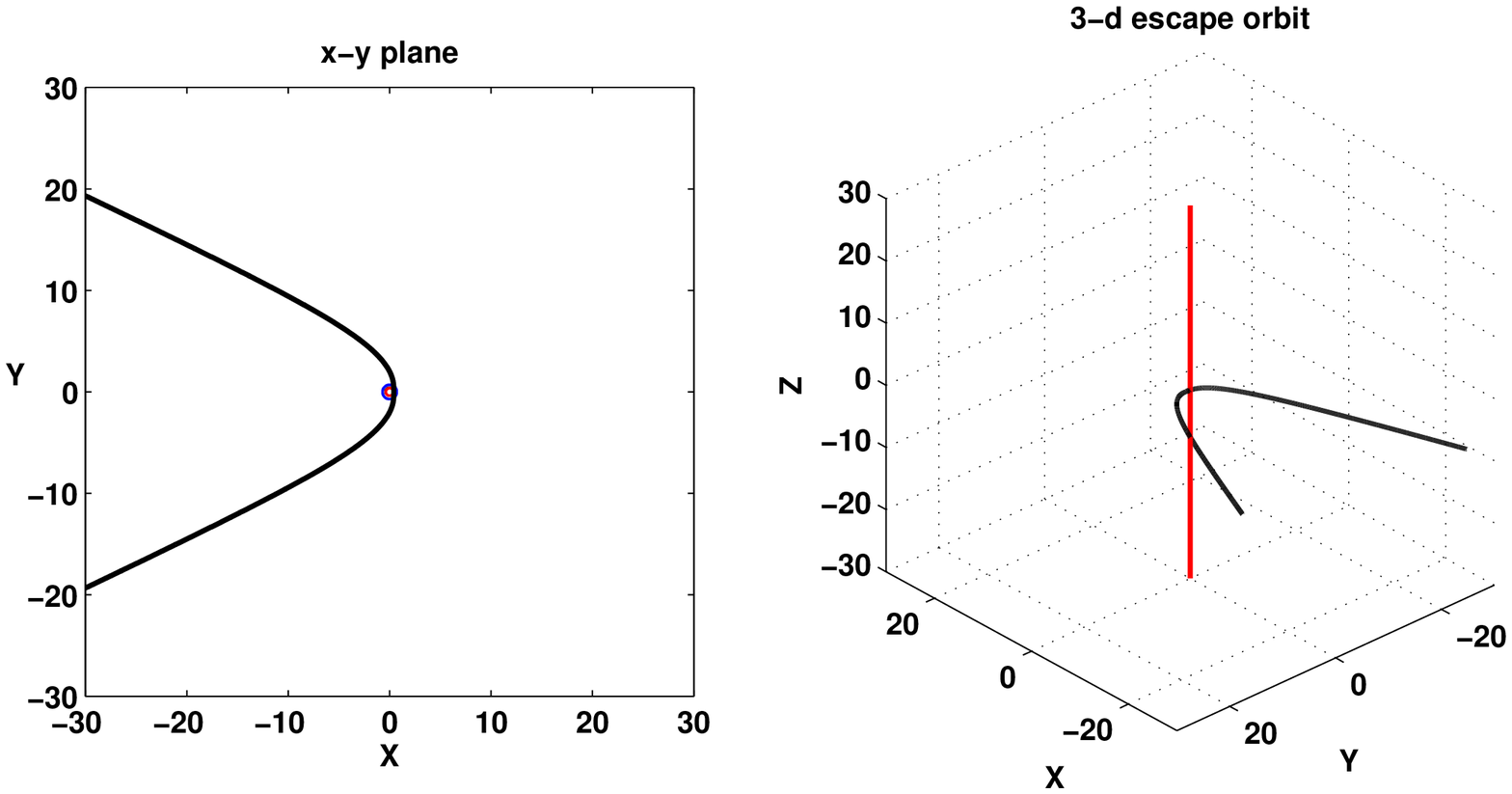}

}
\caption[Optional caption for list of figures]{The change of the escape orbit of
a massive test particle ($\varepsilon=1$) with $E=1.037$, $L_z^2=0.015$ and $p_z=0.03$ 
due to the change of the ratio between Higgs and gauge boson mass $\beta$. Here, we have chosen
$\gamma=0.3$. The red and blue circle indicates the scalar field
core and the magnetic flux tube core of the Abelian--Higgs string, respectively. In the 3--dimensional
plots the red line indicates the string axis.}\label{geo4}
\end{figure}

\subsubsection{Massless particles $\varepsilon=0$}
It seems that massless particles can only move on escape orbits, i.e.
bound orbits as the ones shown for massive particles are not possible.
We show the change of an escape orbit with the gravitational coupling $\gamma$ 
in Fig.\ref{mlEscLzsqrt0030E0075Pz0050lambda050vgamma}.
For sufficiently large $\gamma$ we observe that the
massless test particle encircles the cosmic string before moving again to infinity.
This is completely new as compared to the space--time of an infinitely thin cosmic
string. The change of the orbit with $\beta$ is qualitatively similar to that
of a massive particle. This is why we don't present it here.

\begin{figure}[htp]
\centering

\subfigure[$\gamma = 0.30$]{
\includegraphics[scale=0.5]{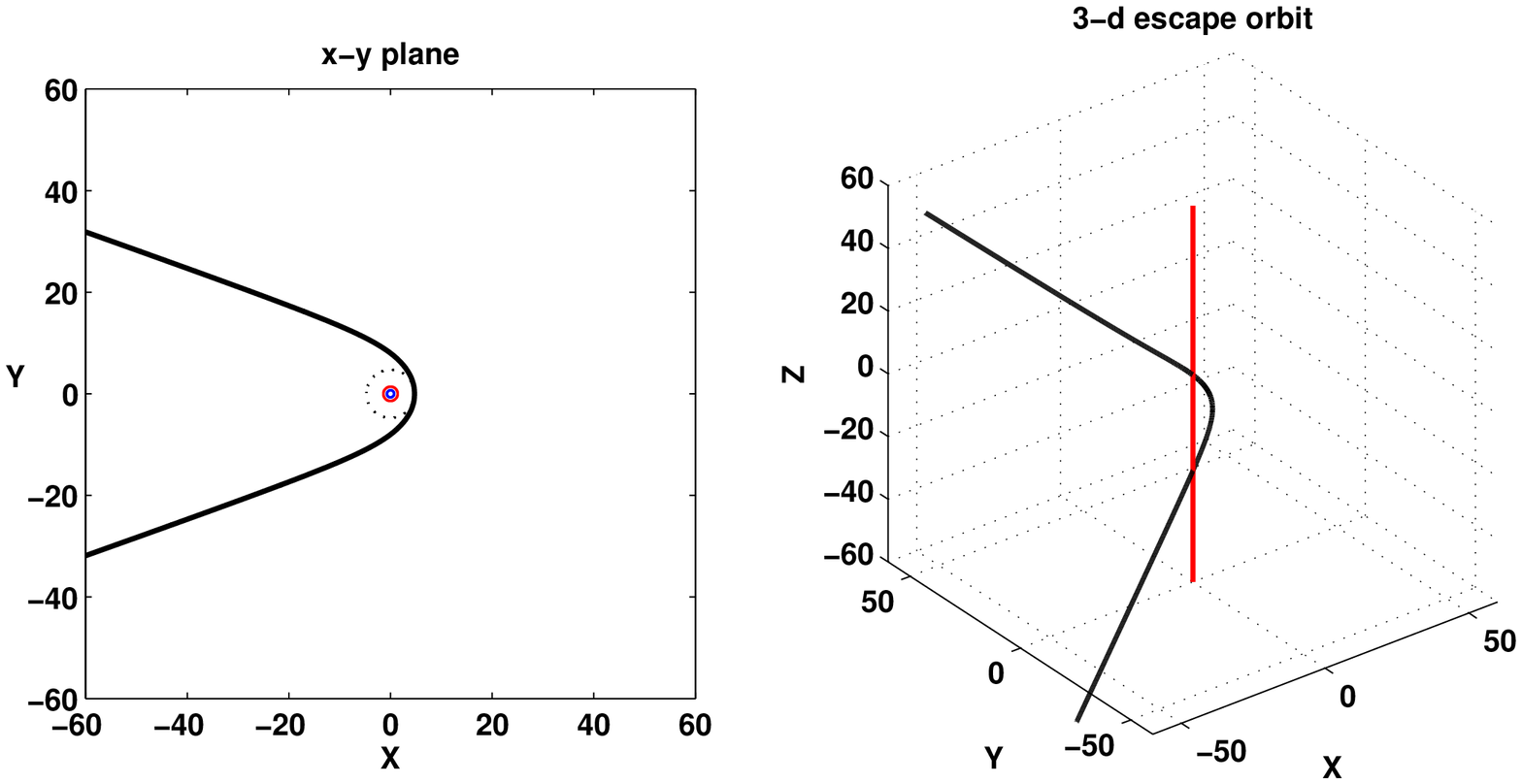}

}

\subfigure[$\gamma = 0.48$]{
\includegraphics[scale=0.5]{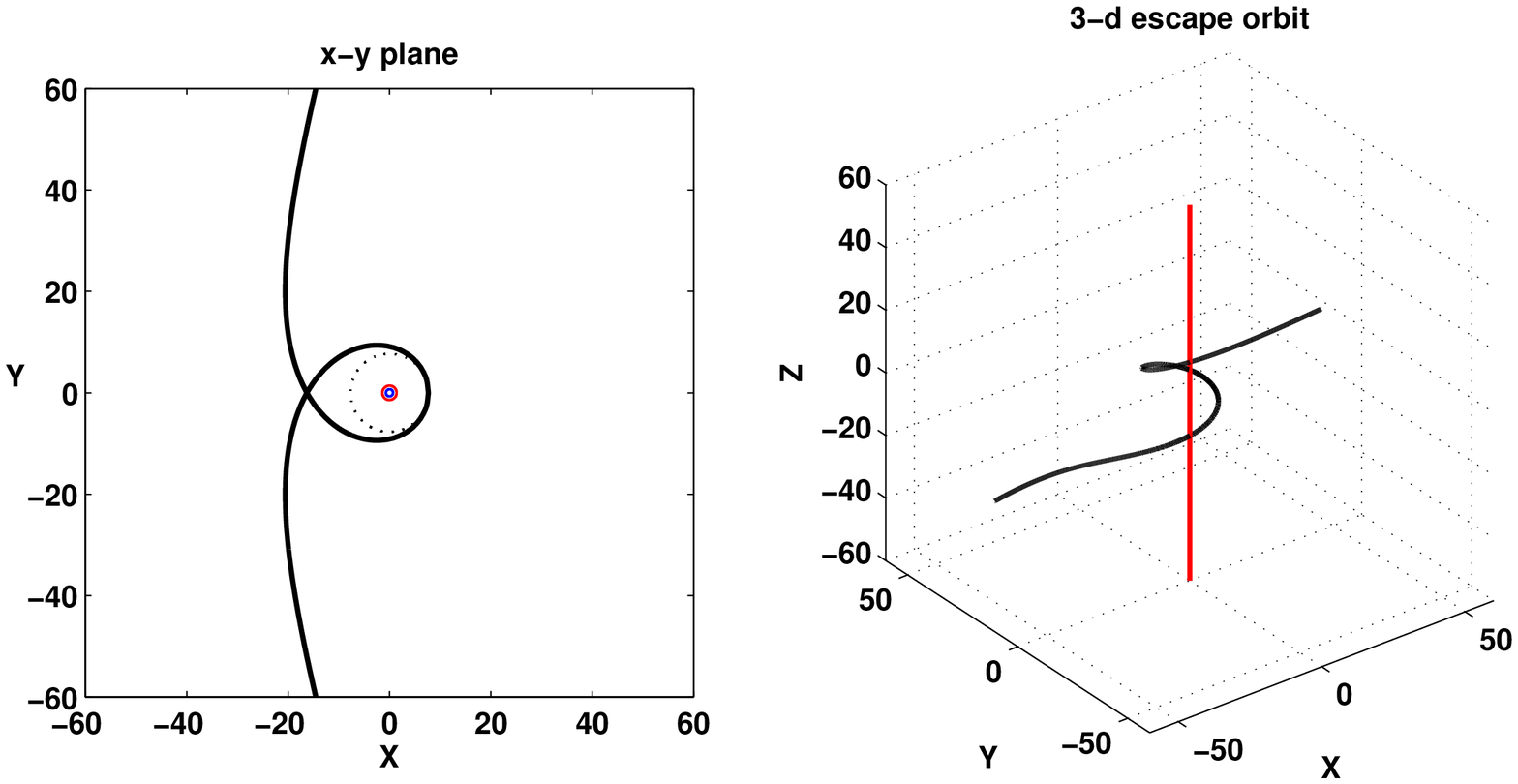}

}
\subfigure[$\gamma = 0.61$]{
\includegraphics[scale=0.5]{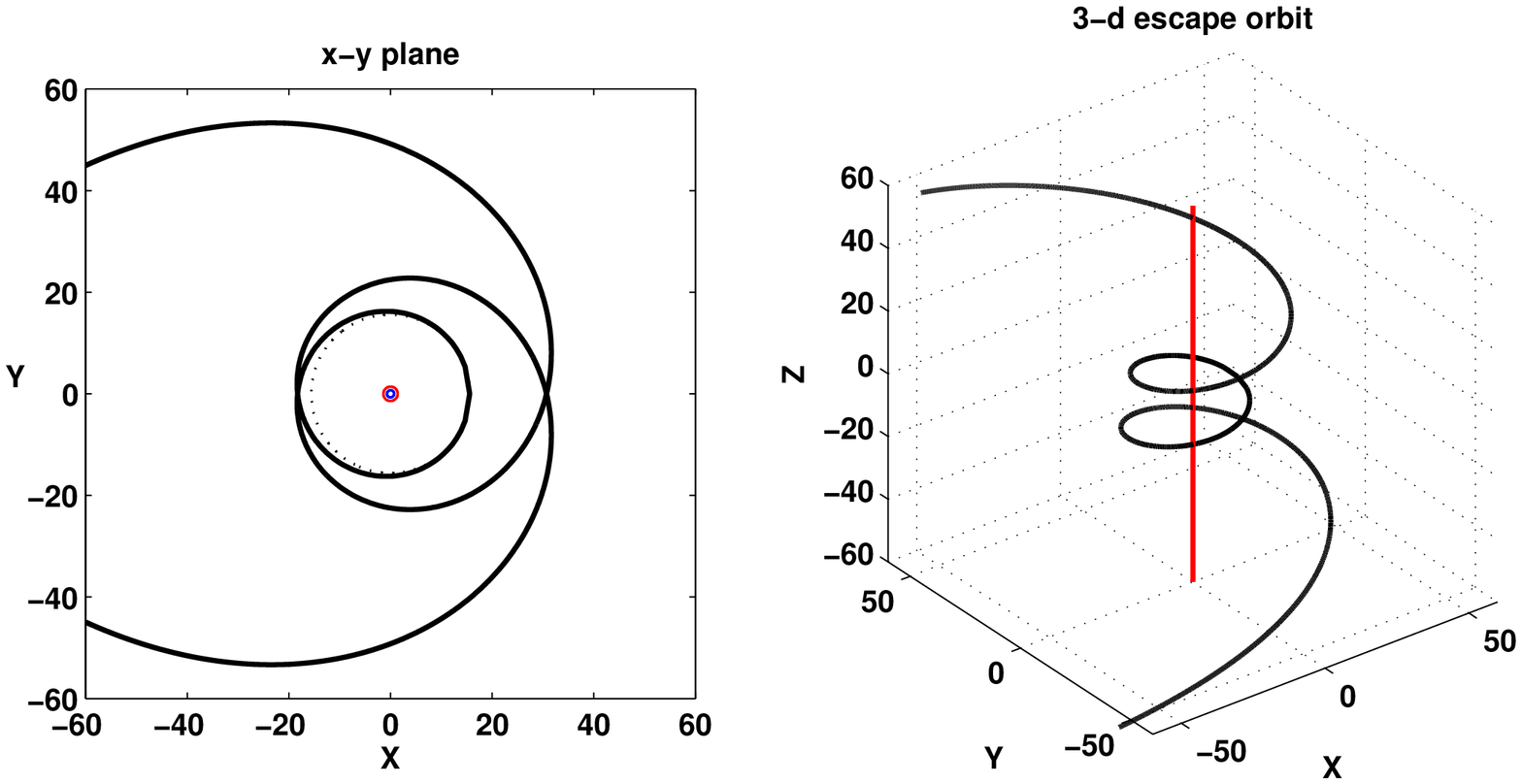}

}

\caption[Optional caption for list of figures]{
The change of the escape orbit of
a massless test particle ($\varepsilon=0$) with $E=0.075$, $L_z^2=0.03$ and $p_z=0.05$ 
due to the change of the gravitational coupling $\gamma$. Here, we have chosen
$\beta=0.5$. The red and blue circle indicates the scalar field
core and the magnetic flux tube core of the Abelian--Higgs string, respectively. In the 3--dimensional
plots the red line indicates the string axis.}\label{mlEscLzsqrt0030E0075Pz0050lambda050vgamma}
\end{figure}

\subsection{Observables}
\subsubsection{Perihelion shift}
\begin{figure}[h!]
\begin{center}
\resizebox{4in}{!}{\includegraphics{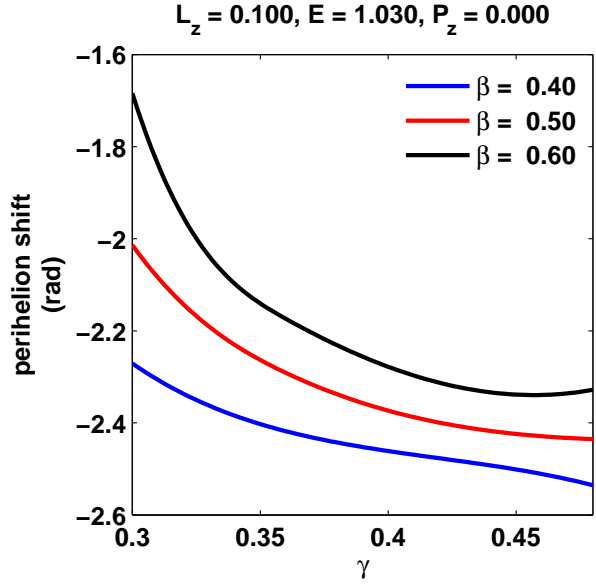}}
\end{center}
\caption{The dependence of the perihelion shift $\delta\varphi$ on the gravitational coupling $\gamma$ for three different ratios $\beta$ of the Higgs to gauge boson mass.
Here, we have chosen $\varepsilon=1$, $L_z =0.1$, $E= 1.03$ and $p_z = 0$.}\label{phLd040506E1031Lz0010Pz00}
\end{figure}

We can calculate the perihelion shift of a planar ($p_z=0$) bound orbit of a massive test
particle ($\varepsilon=1$) by using (\ref{neweq1}). The perihelion shift then reads
\begin{equation}
 \delta \varphi = 2\int\limits_{\rho_{\rm min}}^{\rho_{\rm max}}   \frac{L_z d\rho}{L(\rho)^2\left(\frac{E^2-p_z^2}{N(\rho)^2}-\frac{L_z^2}{L(\rho)^2}-1\right)^{1/2}} - 2\pi  \ ,
\end{equation}
where $\rho_{\rm min}$ and $\rho_{\rm max}$ are the minimal and maximal radius of the bound orbit.
Our results for a particle with $E=1.03$, $L_z=0.1$ and $p_z=0$ are shown in Fig.\ref{phLd040506E1031Lz0010Pz00}, where
the value of $\delta \varphi$ is given in dependence on the gravitational coupling
$\gamma$ for different values of $\beta$. 
For this particular case, the perihelion shift is negative which means that
the particle moves from $\rho_{\rm min}$ to $\rho_{\rm max}$ and back to $\rho_{\rm min}$ under an
angle of less than $2\pi$. Though in most space--times the perihelion shift is positive
it is not too suprising that it can become negative in our case since the space--time is conical.
Typically, the perihelion shift is negative for $\cal{E}$ small, i.e. small values of the energy, while it becomes positive for larger values of $\cal{E}$.

We observe that the value of the perihelion shift decreases for increasing
$\gamma$. Moreover, increasing the ratio $\beta$ between Higgs and gauge boson
mass  increases the value of the perihelion shift.

\subsubsection{Light deflection}

\begin{figure}[h!]
\begin{center}
\resizebox{4in}{!}{\includegraphics{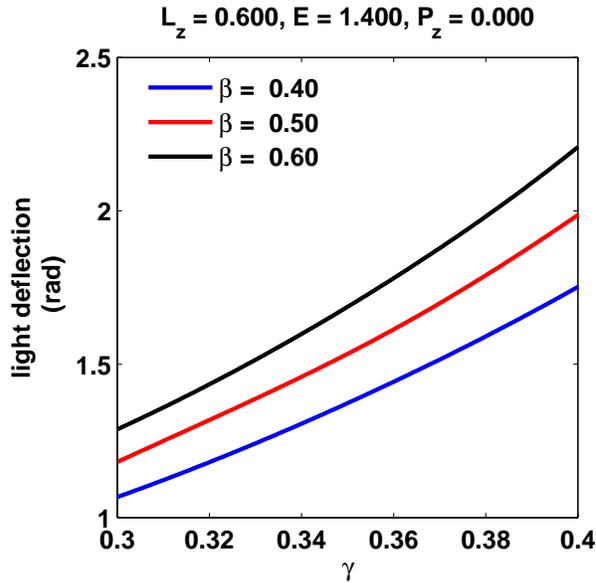}}
\end{center}
\caption{The dependence of the light deflection $\widetilde{\delta\varphi}$ on the gravitational coupling $\gamma$ for three different ratios $\beta$ of the Higgs to gauge boson mass.
Here, we have chosen $\varepsilon=0$, $L_z =0.6$, $E= 1.4$ and $p_z = 0$.
}\label{LiDefLd040506E1400Lz0600Pz00}
\end{figure}

The deflection of light by a cosmic string can be calculated by using
 (\ref{neweq1})  for a planar ($p_z=0$) escape orbit of a massless  test
particle ($\varepsilon=0$). The light deflection then reads
\begin{equation}
 \widetilde{\delta \varphi} = 2\int\limits_{\rho_{\rm min}}^{\infty}   \frac{L_z d\rho}{L(\rho)^2\left(\frac{E^2-p_z^2}{N(\rho)^2}-\frac{L_z^2}{L(\rho)^2}\right)^{1/2}} - \pi  \ ,
\end{equation}
where $\rho_{\rm min}$ is the minimal radius of the orbit.

Our numerical results are shown in Fig.\ref{LiDefLd040506E1400Lz0600Pz00}, where we give
the value of $\widetilde{\delta\varphi}$ in dependence on the gravitational coupling
$\gamma$ for three different values of $\beta$. Apparently, light deflection becomes stronger
when increasing $\gamma$ which results from the increase of the deficit angle. 
Moreover, increasing the ratio between Higgs and gauge boson mass increases the light deflection.

\section{Conclusions}
In this paper, we have studied the geodesic motion of massive and massless test particles
in the space--time of an Abelian--Higgs string. We find that the existence of
bound orbits, i.e. orbits on which a particle moves between a finite maximal
and finite minimal radius depends crucially on the choice of the ratio between
the symmetry breaking scale and the Planck mass and the choice of the ratio between the Higgs and
gauge boson mass. In this paper, we have only considered the case
of small symmetry breaking scale (in comparison to the Planck mass), i.e. we have only
taken space-times with deficit angle smaller than $2\pi$ into account.
Moreover, we have not studied Melvin--type space--times which we believe are physically
not relevant from a astrophysical/cosmological point of view. 

We observe that bound orbits are only possible if the test particle is massive and if the Higgs boson mass is smaller than the gauge
boson mass or in other words if the magnetic flux tube core lies inside the
scalar core of the cosmic string. Increasing either the ratio between the symmetry breaking scale and the
Planck mass or the ratio between the Higgs and gauge boson
mass the particles move closer and closer around the string core. In fact, they move inside
both the scalar and flux tube core and are not restricted to the movement inside
the vacuum region outside both cores. 
For massless particles, which can only move on escape orbits, we observe a new phenomenon as compared
to the space--time of an infinitely thin cosmic string: the particles can encircle the string
before moving again to infinity. 

The perihelion shift of bound orbits of massive particles can be either negative or positive
depending on the particle's energy and decreases with increasing ratio
between symmetry breaking scale and Planck mass. Moreover, it increases  
with increasing Higgs to gauge boson mass ratio. The light deflection by a cosmic
string increases with both the ratio between symmetry breaking scale and Planck mass
and the ratio between Higgs and gauge boson mass.
Since one of the possible detections of cosmic strings would be by light deflection (i.e.
gravitational lensing) our results can be used to compare possible observational
data to make predictions about the symmetry breaking scale at which the cosmic string
formed as well as about the ratio between the corresponding Higgs and gauge boson mass of the underlying field theory. 

It would also be interesting to study geodesic motion in the space--time of a
semilocal string \cite{semilocal,hindmarsh,hartmann_urrestilla}, which is a solution of the electroweak model with
gauge group SU(2)$\times$ U(1) in the limit where the Weinberg angle $\theta_{\rm W}=\pi/2$.
Moreover, field theoretical models describing so--called $p$--$q$--strings have been studied
recently \cite{saffin,hartmann_urrestilla2}. $p$--$q$--strings are  supersymmetric bound states
of $F$-- and $D$--strings and provide one of the examples of strings that might have formed
in inflationary models resulting from string theory. Understanding how test particles
move in the space--time of these strings would provide a further possibility to detect them.

\medskip
\medskip

\noindent
{\bf\large Acknowledgments} 
We thank V. Kagramanova for discussions in the early stages of this project.  The work of PS was supported by DFG grant HA-4426/5-1. We are grateful to  G.~Gibbons for bringing \cite{gibbons}
to our attention.

\newpage

\begin{small}

\end{small}


\begin{thebibliography}{99}

\bibitem{polchinski} see e.g. J. Polchinski, {\it Introduction to cosmic
F- and D-strings}, hep-th/0412244 and references therein.
\bibitem{vs} A. Vilenkin and  P. Shellard, {\it Cosmic strings and other topological defects}, Cambridge University Press (1994).
\bibitem{braneinflation} 
M.~Majumdar and A.~C.~Davis,
  JHEP {\bf 0203} (2002) 056
  [arXiv:hep-th/0202148].
S.~Sarangi and S.~H.~H.~Tye,
  Phys.\ Lett.\  B {\bf 536}, 185 (2002)
  [arXiv:hep-th/0204074].
\bibitem{cmb_cosmic}  
 N.~Bevis, M.~Hindmarsh, M.~Kunz and J.~Urrestilla,
  Phys.\ Rev.\ Lett.\  {\bf 100}, 021301 (2008)
  [arXiv:astro-ph/0702223];
  Phys.\ Rev.\  D {\bf 75}, 065015 (2007)
  [arXiv:astro-ph/0605018]; {\it for a recent review
see} C.~Ringeval, {\it Cosmic strings and their induced non-Gaussianities in the cosmic
  microwave background}, arxiv: 1005.4842 (astro-ph).
\bibitem{khol} {\it see e.g.} M.~Yu.~Khlopov, {\it Cosmoparticle physics}, World Scientific (1999)
and references therein.
\bibitem{gibbons}  G.~W.~Gibbons,
  Phys.\ Lett.\  B {\bf 308}, 237 (1993).
\bibitem{ppm} A.~de Padua, F.~Parisio-Filho and F.~Moraes,
  Phys.\ Lett.\  A {\bf 238} (1998) 153.
\bibitem{ag} 
 A.~N.~Aliev and D.~V.~Galtsov,
  Sov.\ Astron.\ Lett.\  {\bf 14}, 48 (1988).
\bibitem{gm} D.~V.~Galtsov and E.~Masar,
  Class.\ Quant.\ Grav.\  {\bf 6}, 1313 (1989).
\bibitem{cb}  S.~Chakraborty and L.~Biswas,
  Class.\ Quant.\ Grav.\  {\bf 13}, 2153 (1996).

\bibitem{hhls1} E.~Hackmann, B.~Hartmann, C.~Laemmerzahl and P.~Sirimachan,
  Phys.\ Rev.\  D {\bf 81}, 064016 (2010)
  [arXiv:0912.2327 [gr-qc]].

\bibitem{Ozdemir2003} N.~Ozdemir,
  Class.\ Quant.\ Grav.\  {\bf 20} 4409 (2003).
\bibitem{Ozdemir2004}
  F.~Ozdemir, N.~Ozdemir and B.~T.~Kaynak,
  Int.\ J.\ Mod.\ Phys.\  A {\bf 19} 1549 (2004).
\bibitem{Fernandes2006}S.~G.~Fernandes, G.~De A.Marques and V.~B.~Bezerra,
  Class.\ Quant.\ Grav.\  {\bf 23} 7063 (2006).
\bibitem{hhls2} E.~Hackmann, B.~Hartmann, C.~L\"ammerzahl and P.~Sirimachan,
{\it Test particle motion in the space--time of a Kerr black hole pierced by a cosmic
string},  arXiv:1006.1761 [gr-qc].
\bibitem{no} 
 H.~B.~Nielsen and P.~Olesen,
  Nucl.\ Phys.\  B {\bf 61}, 45 (1973).
\bibitem{garfinkle} D.~Garfinkle,
  Phys.\ Rev.\  D {\bf 32}, 1323 (1985).

\bibitem{gl} P.~Laguna and D.~Garfinkle,
  Phys.\ Rev.\  D {\bf 40}, 1011 (1989);
 M.~E.~Ortiz,
  Phys.\ Rev.\  D {\bf 43}, 2521 (1991).

\bibitem{clv} M.~Christensen, A.~L.~Larsen and Y.~Verbin,
  Phys.\ Rev.\  D {\bf 60}, 125012 (1999)
  [arXiv:gr-qc/9904049].

\bibitem{bl}  Y.~Brihaye and M.~Lubo,
  Phys.\ Rev.\  D {\bf 62}, 085004 (2000)
  [arXiv:hep-th/0004043].

\bibitem{kkl} 
V.~Kagramanova, J.~Kunz and C.~Lammerzahl,
  Gen.\ Rel.\ Grav.\  {\bf 40} (2008) 1249
  [arXiv:0708.1747 [gr-qc]].

\bibitem{bps} E.~B.~Bogomolny,
  Sov.\ J.\ Nucl.\ Phys.\  {\bf 24} (1976) 449
  [Yad.\ Fiz.\  {\bf 24} (1976) 861].
\bibitem{colsys} U. Ascher, J. Christiansen and R. Russell, Math. of Comp. {\bf 33}, 659 (1979);
ACM Trans. {\bf 7}, 209 (1981).  

\bibitem{acl} M.~A.~Abramowicz, B~.Carter and J.~P.~Lasota, Gen. Rel. Grav. {\bf 20}, 1173 (1988).
\bibitem{hagi} Y. Hagihara, {\it Theory of relativistic trajectories in a gravitational field of Schwarzschild}, \emph{Japan. J. Astron. Geophys}. \textbf{8}, 67 (1931).
\bibitem{semilocal} 
T. Vachaspati and A. Achucarro, Phys. \ Rev. \ D {\bf 44}, 3067 (1991);
Phys. Rept. {\bf 327}, 347 (2000).

\bibitem{hindmarsh}
M. Hindmarsh,  Phys. Rev. Lett. {\bf 68}, 1263 (1992);
 Nucl. Phys. B {\bf 392}, 461 (1993) [arXiv:hep-ph/9206229].
\bibitem{hartmann_urrestilla} B.~Hartmann and J.~Urrestilla,
  J.\ Phys.\ Conf.\ Ser.\  {\bf 229} (2010) 012008
  [arXiv:0911.3062 [gr-qc]].
\bibitem{saffin}
  P.~M.~Saffin,
  JHEP {\bf 0509}, 011 (2005)
  [arXiv:hep-th/0506138].
\bibitem{hartmann_urrestilla2}
  B.~Hartmann and J.~Urrestilla,
  JHEP {\bf 0807} (2008) 006
  [arXiv:0805.4729 [hep-th]].

\end{thebibliography}
\end{document}